\documentclass[11pt]{article}
\pdfoutput=1
\usepackage{jheppub}
\usepackage{amsmath,amssymb,mathabx,bbm}
\usepackage{enumerate}
\usepackage{xcolor}
\usepackage{graphics}
\usepackage{tikz-feynman}

\newcommand{\Bi}{ \text{Bi} }
\newcommand{\Ai}{ \text{Ai} }
\newcommand{\IBi}{ I_{\Bi} }
\newcommand{\IAi}{ I_{\Ai} }
\newcommand{\M}{ \mathcal{M} }
\newcommand{\E}{ \mathcal{E} }
\newcommand{\rtau}{  \bar{\tau} }
\renewcommand{\b}{  \hat{b} }
\newcommand{\z}{  \hat{z} }
\newcommand{\zz}{  \hat{z} }

\newcommand{\nn}{\nonumber}

\newcommand{\con}{  \bar{c}_{\phi} }
\newcommand{\dis}{  \bar{d}  }

\begin{document}

\title{Novel Screening with Two Bodies: \\ Summing the ladder in disformal scalar-tensor theories}

\author[a]{Anne-Christine Davis}
\author[a, b]{and Scott Melville}
\affiliation[a]{DAMTP, Center for Mathematical Sciences, University of Cambridge, CB3 0WA, UK}
\affiliation[b]{Emmanuel College, University of Cambridge, CB2 3AP, UK}

\emailAdd{a.c.davis@damtp.cam.ac.uk}
\emailAdd{scott.melville@damtp.cam.ac.uk}

\abstract{
\noindent When augmenting our cosmological models or gravitational theories with an additional light scalar field, any coupling between matter and this scalar can affect the orbital motion of binary systems. 
Ordinarily, the new force mediated by the scalar can be naturally the same order of magnitude as the usual gravitational force and therefore is tightly constrained.
We show that a disformal coupling between the scalar and matter can lead to 
a novel screening mechanism in which these fifth forces are suppressed by several orders of magnitude at sufficiently small separations and large relative velocities (such as solar system scales). 
This is a result of resumming a class of ladder diagrams, which suppresses the propagation of scalar signals between the two bodies. 
Moreover, we are able to relate potential ambiguities in this resummation to non-perturbative effects (which are invisible to perturbation theory).  
As a result, solar system tests and future gravitational wave observations can now be used to place meaningful constraints on scalar-tensor theories with disformal couplings. 
We exemplify this using observational bounds on the precession of planetary orbits.   
}


\date{\today}
\maketitle

\section{Introduction}
\label{sec:intro}

Light scalar fields enjoy many uses in late-time cosmology. Theoretically, they represent one of the simplest modifications of gravity beyond General Relativity, introducing only a single new degree of freedom (see e.g. \cite{Capozziello:2007ec, Capozziello:2011et, Clifton:2011jh, Joyce:2014kja, Bull:2015stt, Koyama:2015vza} and references therein). Phenomenologically, they are versatile enough to construct a diverse range of models for the dark sector, frequently employed in studies of dark energy (for instance \cite{ArmendarizPicon:2000dh, Boisseau:2000pr, Copeland:2006wr, Bamba:2012cp}) and dark matter (for instance \cite{Sin:1992bg, Hu:2000ke, Burgess:2000yq, Bekenstein:2004ne}).
Coupling this new light field ($\phi$) to matter generically introduces a long-range \emph{fifth force}. The simplest coupling, a $\phi T_{\mu}^{\;\; \mu}$ conformal interaction, has a long history \cite{Damour:1992we}. It leads to order one corrections to the Newtonian force between two masses and is tightly constrained by fifth force experiments in the solar system \cite{Bertotti:2003rm, Williams:2004qba}---often necessitating some kind of screening mechanism if the model is to be viable \cite{Vainshtein:1972sx, Damour:1994zq, Khoury:2003aq, Babichev:2009ee} (see also \cite{Burrage:2016bwy,Burrage:2017qrf, Sakstein:2018fwz,Baker:2019gxo} for modern reviews).  
In addition, there can also be a disformal coupling of the form $\nabla_\mu \phi \nabla_\nu \phi T^{\mu\nu}$ between matter and a light scalar field, and this has also been constrained via a number of astrophysical and terrestrial experiments \cite{Koivisto:2008ak, Zumalacarregui:2010wj, Koivisto:2012za, vandeBruck:2013yxa, Neveu:2014vua, Sakstein:2014isa, Sakstein:2014aca, Ip:2015qsa, Sakstein:2015jca, vandeBruck:2015ida, vandeBruck:2016cnh, Kaloper:2003yf, Brax:2014vva, Brax:2015hma, Brax:2012ie, vandeBruck:2012vq, Brax:2013nsa}.
Following the advent of gravitational wave \cite{Abbott:2016blz, Abbott:2016nmj} and multi-messenger \cite{TheLIGOScientific:2017qsa, Monitor:2017mdv, GBM:2017lvd} astronomy, the orbits of binary systems will soon also provide a useful probe for these light scalars (e.g. \cite{Wagoner:1970vr, Yunes:2016jcc, Baker:2017hug, Creminelli:2017sry, Sakstein:2017xjx, Ezquiaga:2017ekz, Langlois:2017dyl, Heisenberg:2017qka, Akrami:2018yjz, BeltranJimenez:2018ymu, deRham:2018red, Copeland:2018yuh} and many other recent works).
Providing an accurate and reliable theoretical description of the two-body problem in the presence of a light scalar field has therefore never been more pressing.  

~\\

In this work, we discuss how the motion of a two-body system is affected by a disformal couping between matter and the light scalar field. We find that there are two distinct regimes, controlled by the dimensionless ratio,
\begin{equation}
 L = \frac{  v^2 \sqrt{m_1 m_2} }{ r^3 \, \M^4  }
\end{equation}
which we will call the \emph{ladder parameter}, where $m_A$ are the particle masses\footnote{
For massless particles, $m_A$ is replaced by the particle's energy. 
}, $r$ their typical spatial separation, $v$ their typical relative velocity, and $\M$ is a constant energy scale determined by the disformal coupling strength. 
When $L \ll 1$ (sufficiently large separations and small relative velocities), then the effects of the disformal interaction are small and can be treated perturbatively as an expansion in powers of $L$---this is known as the \emph{ladder expansion} \cite{Brax:2018bow, Brax:2019tcy, Kuntz:2019zef}. Conversely, when $L \gg 1$ (at sufficiently small separations and large relative velocities), the effects of the disformal interaction are large and the ladder expansion is no longer valid. Instead, we show that a \emph{resummation} of the ladder series (to all orders in $L^n$) leads to an efficient \emph{screening} of fifth forces---by over ten orders of magnitude for the planets in our solar system.  

~\\

To illustrate these concepts, we will focus on a simple toy model in which all matter fields ($\psi$) couple to both a spacetime metric ($g_{\mu\nu}$) and a free scalar field ($\phi$) via an effective metric $\tilde{g}_{\mu\nu}$,
\begin{equation}
 S = \int d^4 x \, \sqrt{-g} \left( \frac{M_P^2}{2} R [ g_{\mu\nu} ]  - \frac{1}{2} ( \nabla \phi )^2 \right) + S_m [  \, \psi , \;\; \tilde{g}_{\mu \nu} ( g_{\alpha\beta} , \phi , \nabla_\mu ) \; ] 
\label{eqn:simpleS}
 \end{equation}
We will refer to $g_{\mu\nu}$ as the \emph{Einstein-frame} metric, and $\tilde{g}_{\mu\nu}$ as the \emph{Jordan-frame} metric \cite{Bekenstein:2004ne}. In the absence of the scalar field, test particles would move along geodesics of $g_{\mu\nu}$ and would experience the usual Newtonian force with universal strength $G_N = 1/ (8 \pi M_P^2 )  =  6.71 \times 10^{-57} \text{eV}^{-2}$ in the non-relativistic limit. 
Once $\phi$ is turned on, test particles will instead follow geodesics of the Jordan-frame $\tilde{g}_{\mu\nu}$, and hence experience a new apparent force.

For the simple Jordan-frame metric, 
\begin{equation}
 \tilde{g}_{\mu \nu} ( g_{\alpha \beta} , \phi , \nabla_\mu ) = g_{\mu \nu} +  \con \frac{\phi}{M_P} g_{\mu\nu} + \dis  \frac{ \nabla_{\mu} \phi \nabla_\nu \phi   }{ \M^4 }  \, ,
\end{equation}
this scalar-mediated fifth force has two components: a \emph{conformal} piece controlled by the $\con \, \phi \, g_{\mu\nu}$ term, with typical strength comparable to $G_N$ (for $\con \sim \mathcal{O} (1)$ an order one dimensionless coupling), and a \emph{disformal} piece controlled by the $\dis \, \nabla_\mu \phi \nabla_\nu \phi$ term, with typical strength set by the constant scale $\M$ (for $\dis \sim \mathcal{O} (1)$ an order one dimensionless coupling). 
The ladder expansion treats this disformal term as a small perturbation, but will break down once the disformal coupling competes with the conformal coupling to matter, 
\begin{equation}
\text{Ladder expansion breaks down when:} \;\;\;\; \frac{ \nabla_\mu \phi \nabla_\nu \phi }{ \M^4} 
\gtrsim \frac{\phi}{M_P} \, g_{\mu\nu}
\label{eqn:intro_breakdown}
\end{equation}
which naively happens when $(\nabla \phi)^2/\phi \sim \nabla^2 \phi \gtrsim \M^4/M_P$. 
The condition \eqref{eqn:intro_breakdown} when moving past a point source with velocity $v$ corresponds to $L \gtrsim 1$.   

If $\phi$ were to play the role of dark energy\footnote{
We stress that we focus on the simple toy model \eqref{eqn:simpleS} to highlight the effects of a disformal coupling, and it should not be viewed as a viable theory for dark energy---the disformal effects discussed here are sufficiently general that in future they may be embedded in a more sophisticated scalar-tensor theory in which $\phi$ is a viable dark energy candidate. 
}, one typically requires $1/\M^2 \sim 1/(M_P H_0) \sim 10^6 \, \text{eV}^{-2}$ to reproduce $\Lambda$CDM-like background evolution, and this is over 60 orders of magnitude stronger than $G_N$. As a result, it is very easy for the disformal piece to be the dominant contribution to the fifth force. In particular, for a planet of mass $m$ on a gravitationally bound orbit around the Sun ($v^2 \sim G_N M_{\rm Sun} / r $), 
the ladder parameter is given by,  
\begin{equation}
 L \approx  10^{16}  \left( \frac{ \text{1 AU} }{ r} \right)^4  \left(  \frac{ m }{  M_{\rm Sun}  } \right)^{1/2}
\end{equation}   
and the ladder expansion is only valid for Earth-sized planets if the orbit is larger than $r \sim 10^{3} $ AU! It is therefore vital, if attempting to constrain disformally coupled dark energy using planetary motion in our solar system, to go beyond the perturbative ladder expansion and capture the $L \gg 1$ regime of this theory.   
 
~\\

In the following, we will show how to successfully resum the ladder expansion to all orders in $L$, providing solutions to the classical equations of motion which are valid at any separation and any relative velocity. 
When taking the limit $L \gg 1$, we find that the disformal interaction results in an efficient screening of the field (rather than producing very large effects, which might have been naively expected).  
A complementary way to view this resummation is as a modification to the $\phi$ propagator: the disformal coupling plays the role of a new effective kinetic term for the scalar, and in regions where $L \gg 1$ this modification suppresses the propagation of $\phi$ signals between the two bodies.  

Interestingly, we find that this resummation is not always unique: there is an ambiguity associated with the presence of non-perturbative effects. The perturbative resummation is providing novel information about non-perturbative physics, an unexpected example of \emph{resurgence} (see  \cite{Dorigoni:2014hea} and references therein). 
This resurgence takes place classically, but can nonetheless be viewed as an instanton contributions arising from different saddle points in the effective action which determines $\phi [ x(\tau) ]$ as a worldline functional.    
 
Finally, given the similarity in scales and the breakdown condition \eqref{eqn:intro_breakdown}, it is tempting to regard this ladder screening as some Einstein-frame manifestation of the usual Vainshtein mechanism \cite{Vainshtein:1972sx}.  
If there were such a connection, it would be very remarkable: since Vainshtein relies on non-linearities, computing the screened profile around multiple bodies is a long-standing open problem. 
However, the ladder screening observed here differs from Vainshtein in a number of aspects. For instance, it does not seem to occur for single bodies, is absent if both particles are at rest, and is controlled by the geometric mean $\sqrt{m_1 m_2}$ of the masses. 
We will show explicitly in an upcoming work \cite{us} that an explicit field redefinition of the simple model \eqref{eqn:simpleS} to the Jordan-frame results in a particular Horndeski theory (namely the quartic DBI Galileon \cite{deRham:2010eu, Zumalacarregui:2012us}) in which the gravitational mixing \emph{exactly cancels} the Galileon-type operators which are usually responsible for Vainshtein. Consequently, this ladder screening appears to be independent of the Vainshtein mechanism. 
Furthermore, note that our screening is also distinct from the ``disformal screening'' mechanism proposed in \cite{Zumalacarregui:2012us, Koivisto:2012za} (in which large $\rho$ means that $\phi$ is independent of the local energy density), the ``derivative chameleon'' of \cite{Noller:2012sv} (see also the discussion in \cite{Sakstein:2014isa}), and other kinetic screening mechanisms in which $X := (\nabla \phi)^2$ becomes large (here, $\phi/M_P$ and $(\nabla \phi)^2/\M^4$ both remain small on all scales of interest). 
As far as we are aware, the ladder resummation presented here results in a novel, genuinely two-body,  screening effect. 

~\\

We will now turn to Section~\ref{sec:VV} and provide a more general discussion of the scales at which a perturbative ladder expansion is valid, and then in Section~\ref{sec:2body} return to the simple theory \eqref{eqn:simpleS} and show explicitly how to go beyond the ladder expansion for binary systems (with $S_m$ being two point particles). This generically results in up to four undetermined constants which are not fixed by the ladder expansion. 
In Section~\ref{sec:eg}, we will solve (at leading order in a Post-Newtonian expansion) the resummed ladder equation for the scalar field and show that the resulting fifth forces are screened like $1/L$ in the $L \gg 1$ regime.
We will also demonstrate explicitly that the undetermined constants can be viewed as ambiguities in the Borel resummation of the ladder series, which are often attributed to instanton-like (non-perturbative) effects.   
Finally, in Section~\ref{sec:solarsystem} we compute the backreaction of these fifth forces on an elliptic Keplerian orbit, which leads to an additional precession of the perihelion. Using observational bounds on the precession of planets in our solar system, we place constraints on the conformal/disformal couplings in the toy model \eqref{eqn:simpleS} as a simple demonstration of how the screening opens up large regions of parameter space which may have otherwise been (incorrectly) excluded on the basis of the ladder expansion (which is not valid for low $\M$, for instance the scales encountered in dark energy).   
Finally, we conclude in Section~\ref{sec:disc} with some discussion of future work.

\paragraph{Conventions:}
Greek indices ($\mu,\nu,...$) denote spacetime coordinates $(0,1,2,3)$, while lowercase Latin indices ($i,j,...$) denote spatial coordinates $(1,2,3)$. Uppercase Latin indices ($A,B,...$) denote the particle label (1 or 2), and the variable $\nu$ will denote the mass ratio $m_2/m_1$.  
Covariant derivatives, $\nabla_\mu$, are defined with respect to $g_{\mu \nu}$, which is used to raise/lower indices, and also to normalize delta functions, $\int d^4 x \, \sqrt{-g} \delta^{4} (x) = 1 $. 
The first derivative of $\phi$ can be used to form the Lorentz scalar $X= g^{\mu\nu} \nabla_\mu \phi \nabla_\nu \phi$ in the Einstein frame. An over-tilde denotes analogous tensor quantities constructed with respect to the effective $\tilde{g}_{\mu \nu}$, e.g. $\tilde{\nabla}, \, \tilde{X}, \, $ etc. 
The speed of light\footnote{
We will not introduce photons explicitly, but of course they will have different apparent speeds if measured with respect to $g_{\mu\nu}$ versus $\tilde{g}_{\mu\nu}$. By ``speed of light'' here, we simply mean that coordinates are chosen so that a Minkowski background metric for $g_{\mu\nu}$ has $-\eta_{00} = \eta_{11} = \eta_{22} = \eta_{33}$. 
} will be set to one throughout, but the Planck mass, $M_P$, will always be explicitly included. The Minkowski metric is given by $\eta_{\mu\nu} = \text{diag} (-1, +1,+1,+1)$ in Cartesian coordinates.

\section{Breakdown of the Ladder Expansion}
\label{sec:VV}
We will focus on metric theories of gravity, $g_{\mu\nu}$, plus an additional scalar field, $\phi$, (playing the role of a dark sector), coupled to external matter sources (denoted $\psi$),
\begin{equation}
S = S_{ST} [ g_{\mu \nu} , \phi , \nabla_\mu  ] + S_m [ \tilde{g}_{\mu \nu} ,  \psi , \tilde{\nabla}_\mu ] \, .
\label{eqn:action}
\end{equation}

The classical equations of motion are,
\begin{align}
-2 \mathcal{E}_{g}^{\mu \nu} &=  \frac{\delta \tilde{g}_{\alpha \beta}}{\delta g_{\mu \nu}} T^{\alpha \beta}   \,  ,  \label{eqn:geom} \\
-2 \mathcal{E}_{\phi}  &=  \frac{\delta \tilde{g}_{\alpha \beta} }{\delta \phi} T^{\alpha \beta}  \, , \label{eqn:peom}
\end{align}
where the vacuum equations are $\mathcal{E}_g^{\mu \nu} = \delta \mathcal{L}_{ST} / \delta g_{\mu \nu} $ and $\mathcal{E}_{\phi} = \delta \mathcal{L}_{ST} / \delta \phi$, and the energy-momentum tensor is\footnote{
Note that this $T^{\mu \nu}$ differs from the Jordan frame $\tilde{T}^{\mu\nu}$ by a volume factor, $\sqrt{-\tilde{g}}/\sqrt{-g}$.
},
\begin{equation}
T^{\mu \nu} =  \frac{2}{\sqrt{- g }} \frac{\delta S_m}{\delta \tilde{g}_{\mu \nu} }  \;  . 
\end{equation}

\paragraph{Bekenstein Metric:}
Bekenstein argued that the most general coupling compatible with causality may have a conformal and a disformal part \cite{Bekenstein:1992pj},
\begin{equation}
\tilde{g}_{\mu \nu} = C ( \phi , X ) g_{\mu \nu} + D ( \phi, X ) \nabla_\mu \phi \nabla_\nu \phi \, ,
\end{equation}
where $X = \nabla_\mu \phi \nabla^\mu \phi$. This gives equations of motion,
\begin{align}
-2 \mathcal{E}_g^{\mu \nu}  &=  C  T^{\mu \nu}     +  T_{\phi}^{\mu \nu}  + \nabla^\mu \phi \nabla^\nu \phi \left(  C_{,X}  T + D_{,X}  \nabla_{\alpha} \phi \nabla_{\beta} \phi T^{\alpha \beta} \right)     \\ 
-2 \mathcal{E}_{\phi}  &=  C_{,\phi} \, T  + D_{,\phi}  \, T^{\mu \nu}  \nabla_\mu \phi \nabla_\nu \phi   \nonumber  \\
&- 2 \nabla_\alpha \left[  D \, T^{\alpha \beta} \nabla_\beta \phi  +  \nabla^\alpha \phi  \left( C_{,X} g_{\mu \nu} + D_{,X} \nabla_\mu \phi \nabla_\nu \phi \right) T^{\mu \nu} \right] \, , 
\end{align}
where $T = T^{\mu \nu} g_{\mu \nu}$ is the trace of the energy-momentum tensor. The choice $C=1, D=0$ recovers GR (the scalar field decouples), but generally the conformal and disformal couplings provide new source terms for the fields. 

\paragraph{Power Counting Scales:}
We are going to treat \eqref{eqn:action} as a low-energy effective field theory (EFT). In particular, we are going to expand in the dimensionless quantities,
\begin{equation}
\frac{\phi}{M_P} , \;\; \frac{\nabla \phi}{ \M^2 }  \;\;\;  \ll \;\; 1 
\label{eqn:power_counting}
\end{equation}
where for dark energy one often chooses $\M^2 = M_P H_0$, where $H_0$ is the Hubble rate today (but we will endeavour to keep $\M$ as general as possible, so our analysis also applies to phenomenological dark matter models or strong gravity corrections from some underlying UV theory).
For instance, the Bekenstein metric should be read as,
\begin{equation}
\tilde{g}_{\mu \nu} = \hat C \left(  \frac{\phi}{M_P} , \frac{X}{\M^4} \right) g_{\mu \nu} + \hat D \left(  \frac{\phi}{M_P} , \frac{X}{\M^4} \right) \frac{\nabla_\mu \phi \nabla_\nu \phi}{\M^4} \, ,
\end{equation}
where $\hat{C}$ and $\hat{D}$ are dimensionless functions.

\paragraph{Flat Background:}
Consider expanding around a flat background,
\begin{equation}
g_{\mu \nu} = \eta_{\mu \nu} + \frac{h_{\mu \nu}}{M_P} , \;\;\;\; \phi = 0 + \varphi
\end{equation}
where the fluctuations $(h_{\mu \nu}, \varphi)$ and their derivatives $( \nabla_\alpha h_{\mu \nu}, \, \nabla_\alpha  \varphi)$ are smaller than $M_P$ and $\M^2$ respectively. 
To leading order in $M_P$ and $\M^4$, the Bekenstein metric can then be expanded\footnote{
Note that some authors include an additional factor of 2  in the conformal coupling ($2 \con \varphi / M_P$) so that the $\varphi$ sourced by a point mass is equal to $\con \, \Phi_N$, the Newtonian potential. However, then the effective one-body metric has awkward numerical factors. We prefer this normalisation ($\con \varphi/M_P$) in which $\varphi = \con \Phi_N/2$ around a point mass, since the force communicated by both polarisations of the graviton should physically be a factor of two larger than the force communicated by a single scalar (for order one couplings).  
},
\begin{equation}
\tilde{g}_{\mu \nu} = \eta_{\mu \nu} + \frac{1}{M_P} h_{\mu\nu}  + \con  \frac{\varphi}{M_P} \eta_{\mu \nu}  + \dis \frac{ \partial_\mu \varphi \partial_\mu \varphi}{\M^4} + ... 
\end{equation}
where $\con$ and $\dis$ are now \emph{constants} (and we have rescaled\footnote{
In principle one could also rescale $\M$ to set $| \dis | =1$ at this order, but to keep track of the overall \emph{sign} of the disformal interaction $\dis$ should not be removed completely. We will instead fix $\M$ according to the desired physics of $\phi$ (whether it is dark matter, dark energy, etc.), and then consider $\dis$ as a free parameter which takes values $\sim \mathcal{O} (1)$ of either sign.   
} so that $C( 0, 0) = 1$). Note that there is, in principle, an additional term at this order from the conformal interaction, $ ( \nabla \varphi )^2 \eta_{\mu \nu} $, but we will not include that here\footnote{
Note that since both $T$ and $\Box \varphi$ have support only on the worldlines (see \eqref{eqn:phieom}), the product $T \Box \varphi$ will not contribute to our equation of motion for $\varphi_A$. This $(\nabla \varphi)^2 T$ term therefore only contributes a term like $\nabla_\mu \varphi \nabla^\mu T$, but since $\nabla^\mu T$ scales like an acceleration it is often a small effect. 
}.  

For now, let us assume that $\mathcal{E}_\phi$ is linear in the field---i.e. if the scalar-tensor $S_{ST}$ has a canonical kinetic term for $\phi$, then the homogeneous equation of motion is simply,
\begin{equation}
\mathcal{E}_{\phi} = \Box \varphi + ... 
\end{equation}
where $\Box = \eta^{\mu \nu} \partial_\mu \partial_\nu$. 
We will return to possible non-linearities in $\mathcal{E}_\phi$ at the end of this section. 
Similarly, a canonical kinetic term for the metric gives the usual Einsten tensor\footnote{
$\mathcal{E}_g^{\mu\nu}$ also contains the energy-momentum from $\phi$'s kinetic term, $ \nabla^\mu \phi \nabla^\nu \phi - \tfrac{1}{2} g^{\mu\nu} X$, but this is subleading in the metric equation \eqref{eqn:heomT}. 
},
\begin{equation}
\mathcal{E}_g^{\mu \nu}  = M_P^2 \left( R^{\mu \nu} - \frac{1}{2} R g^{\mu \nu} \right) \, . 
\end{equation}

The metric perturbations in de Donder gauge\footnote{
The linearised gauge condition sets $ \nabla^{\beta} h_{\beta \alpha} = \frac{1}{2} \nabla_\alpha h $, i.e. the trace-reversed graviton is transverse. 
} then obey the usual linearised Einstein equation, namely the expansion of \eqref{eqn:geom}, 
\begin{equation}
\Box h^{\mu\nu} - \frac{1}{2} \eta^{\mu\nu} \Box h  = \frac{1}{M_P} T^{\mu \nu}  
+ ... 
\label{eqn:heomT}
\end{equation}
while the scalar field equation \eqref{eqn:peom} is now sourced by just two terms,
\begin{equation}
\Box \varphi = - \frac{\con}{2 M_P} T   +  \frac{\dis}{\M^4}  \nabla_\mu \left[ \nabla_\nu \varphi  \, T^{\mu \nu}  \right] + ... 
\label{eqn:phieom}
\end{equation} 
where $T = T^{\mu \nu} g_{\mu \nu}$ is the Einstein-frame trace of the energy-momentum tensor. 
Note that the disformal coupling between $\varphi$ and $T_{\mu \nu}$ is \emph{linear} in the scalar field, and so for a given $T^{\mu \nu}$ any superposition of solutions ($\varphi_1 + \varphi_2$) is also a solution.

\paragraph{Ladder Expansion:}
When faced with a source term like $\varphi T$ in \eqref{eqn:phieom}, it is tempting to treat it perturbatively. That is to say, if one expands the field,
\begin{equation}
\varphi =  \varphi^{(0)} +  \varphi^{(1)} + ... 
\label{eqn:varphiexp}
\end{equation}
such that the leading order piece solves,
\begin{equation}
\Box \varphi^{(0)}  =  - \frac{ \con }{2 M_P} T \, ,
\label{eqn:varphi0}
\end{equation}
then one can study the effects of the $\varphi T$ mixing as a small correction.
\begin{equation}
\Box \varphi^{(1)}  =  \frac{ \dis }{\M^4} \,  \nabla_\mu \left[ \nabla_\nu \varphi^{(0)}  \, T^{\mu \nu}  \right] \, .
\label{eqn:varphi1full}
\end{equation}
This provides a way of solving for the field profile perturbatively\footnote{
Rather than treating the disformal term as an ``interaction'' for $\varphi$, one could alternatively view it as a modification to the propagator. For point particles, $T^{\mu\nu}$ is singular and so it is more useful to treat it as an interaction to be resummed. 
}, to any desired order in $1/\M^4$. 
Each term in this expansion can be represented as a \emph{ladder diagram}, in which the scalar field sourced by the matter distribution at leading order then acts as a new effective coupling which generates subleading corrections to the scalar field. For the case of two point particles, the diagrams corresponding to $\varphi^{(0)}$, $\varphi^{(1)}$ and $\varphi^{(2)}$ are shown in Figure~\ref{fig:resum}.

\paragraph{Validity of the Ladder Expansion:}
When does this expansion break down? To estimate the relevant scales at which this expansion is trustworthy, let us assume that $T^{\mu \nu}$ describes a localised mass $m$ a spatial distance $r$ away,
\begin{equation}
T \sim \frac{m}{r^3} \;\;\;\; \Rightarrow \;\;\;\; \varphi^{(0)} \sim \frac{m}{M_P r} \, .
\end{equation}
The first correction, $\varphi^{(1)}$, can then be estimated from \eqref{eqn:varphi1full} as,
\begin{equation}
\varphi^{(1)} \sim  \frac{m^2}{M_P \M^4 r^4}   \;\;\;\; \Rightarrow \;\;\;\; \frac{\varphi^{(1)}}{\varphi^{(0)}} \sim  \frac{m}{\M^4 r^3} \, .
\end{equation}
The corrections from subsequent ladder diagrams can become order one when,
\begin{equation}
r^3 \sim \frac{m}{\M^4} =: R_V^3 \, .
\end{equation}
This is also the scale at which higher derivative terms can become important in theories with (weakly broken) Galileon symmetry, and is conventionally referred to as the \emph{Vainshtein radius}. We will return to these additional operators at the end of this section---here, we simply wish to emphasis that {\bf at distances smaller than $R_V$, the perturbative ladder expansion may no longer be a good approximation}.  

We stress that these rough scaling arguments can over-estimate the size of corrections, as we have assumed that all derivatives $\partial_\mu \sim 1/r$. This is, of course, not quite the case---for instance in static or spherically symmetric situations, then time and angular derivatives are hugely suppressed. A more refined estimate of the scale at which the ladder approximation breaks down is given by considering $T^{\mu \nu} \sim \frac{m}{r^3} u^\mu u^\nu$, where $u^\mu$ is the four-velocity of the mass. This then suggests that,
\begin{equation}
\frac{\varphi^{(1)}}{\varphi^{(0)}} \sim   \frac{ m v^2}{ \M^4 r^3} \, , 
\end{equation} 
and the scale at which the ladder approximation breaks down is rather $r^3 \sim v^2 R_V^3$, a factor of $v^2$ \emph{smaller} than the conventional Vainshtein radius. We will refer to this dimensionless combination as the \emph{ladder parameter}\footnote{
In two-body systems, the relevant Vainshtein scale is the geometric mean, $\sqrt{ R_{V_1} R_{V_2} }$, as we will show explicitly in Section~\ref{sec:2body}. 
},
\begin{equation}
L \sim \frac{ v^2 R_V^3 }{ r^3}  \, ,
\end{equation}
such that the ladder expansion is no longer valid whenever $L \gtrsim 1$. This scale was also recently observed in \cite{Kuntz:2019zef}.  
 
\paragraph{The Need for Resummation:}
The key observation that we wish to make is that perturbative expansions of the kind above are only valid at sufficiently \emph{large distances} and sufficiently \emph{low velocities} (i.e. when $L \ll 1$). 
For instance, if $\phi$ were to play the role of dark energy on cosmological scales, the typical scale required for $\M^2$ ($\sim M_P H_0$)  would fix $R_V \sim 10^8$ AU around the Sun and $R_V \sim 10^6$ AU around the Earth. Consequently, if one imagines placing an Earth-like mass at a distance of 1 AU from the Sun, the condition $L \ll 1$ corresponds to $v^2 \ll 10^{-21}$ (in units where the speed of light is unity). Even the smallest of relative motions with respect to the Sun, at an Earth-like orbital distance, would place us firmly in the regime $v^2 R_V^3 \gg r^3$ and invalidate a ladder expansion.  

Furthermore, for gravitationally bound binary systems, the typical virial velocity is $v^2 \sim R_S/r$, where $R_S$ is the Schwarzschild radius (of $m_1+m_2$). The ladder expansion for dark energy ($\M^2 \sim M_P H_0$) would then require $H_0^2 r^2 \gg R_S^2/r^2$, where $H_0$ corresponds to the Hubble rate today. The ladder expansion breaks down in this case whenever the orbital size in Hubble units becomes comparable to the inverse orbital size in Schwarzschild radii (a useful rule of thumb).  

Rather than describing the effects of the disformal coupling perturbatively, what is required at short distances is a \emph{resummation} of the ladder diagrams, $\sum_{n=0}^{\infty} \varphi^{(n)}$. This is possible in simple cases, as we will show for two point particles in Section~\ref{sec:2body} (and depicted graphically in Figure~\ref{fig:resum}).

\paragraph{Comparison with Vainshtein Mechanism:}
While the $\varphi T$ mixing is linear in $\varphi$, if treated perturbatively as an interaction we have seen that it leads to a cutoff of order $R_V$. 
Contrast this with the genuine non-linearities which may appear in $\mathcal{E}_{\phi}$---for instance, a cubic Galileon interaction in $S_{ST}$ would give,
\begin{equation}
\mathcal{E}_{\phi} = \Box \varphi +  \frac{ g_{3} }{\Lambda_3^3} \nabla_\mu \nabla^{[\mu} \varphi  \nabla_\nu \nabla^{\nu]} \varphi  + ... 
\label{eqn:cubicGal}
\end{equation}
where $\Lambda_3$ is a new constant scale which suppresses higher derivative operators, and can be as low as $\M^4 / M_P$ (thanks to the Galileon symmetry which is only weakly broken by gravitational interactions \cite{Pirtskhalava:2015nla}). 
In this case, if one solves \eqref{eqn:varphi0} for $\varphi^{(0)}$, then the $\varphi^{(1)}$ correction depends on both the non-linearities in $\mathcal{E}_\phi$ and the matter coupling,
\begin{equation}
\Box \varphi^{(1)}  =   \frac{ \dis }{\M^4} \, \nabla_\mu \left[ \nabla_\nu \varphi^{(0)}  \, T^{\mu \nu}  \right] - \frac{ g_{3} }{\Lambda_3^3} \nabla_\mu \nabla^{[\mu} \varphi^{(0)}  \nabla_\nu \nabla^{\nu]} \varphi^{(0)} \, . 
\end{equation}
A rough scaling argument as before indicates that,
\begin{equation}
\frac{\varphi^{(1)}}{\varphi^{(0)}} \sim  \frac{R_V^3}{r^3} \left[  \dis \,  v^2  +  g_3   \frac{\M^4}{M_P \Lambda_3^3}  \right] .
\end{equation}
Note that in the weakly broken Galileon power counting, $M_P \Lambda_3^3 \sim \M^4$, and so $R_V$ is also the scale at which non-linearities from $\mathcal{E}_{\phi}$ will become important. 
One consequence of this is that the effects of the disformal coupling (for non-relativistic velocities, $\dis v^2 \ll g_3$) are always \emph{subdominant} to the higher derivative operators, if the latter have the lowest possible scale which is protected by the broken Galileon symmetry. 

In this work, we will focus on low-energy theories which do \emph{not} have a parametric separation of scales between $\Lambda_3$ and $\M$. For instance, rather than integrating out new (Galileon-invariant) physics at the scale $\Lambda_3$ (which would produce e.g. \eqref{eqn:cubicGal}, with $\M^4 \sim \Lambda_3^3 M_P$), we instead imagine that heavy fields have been integrated out directly at the scale $\M$, where the only symmetry is an approximate shift symmetry for $\phi$. 
In an upcoming work \cite{us}, we demonstrate that this choice is also consistent with the physics in the Jordan-frame: a field redefinition of a purely disformal coupling in the Einstein-frame (which lives at $\M$) into the Jordan-frame results in a special Horndeski theory (which enjoys a cancellation of every operator below $\M$) that does not exhibit conventional Vainshtein screening at $R_V$. 

We will now focus on the simple action \eqref{eqn:simpleS} for a free scalar field in the Einstein frame, with constant conformal and disformal couplings to matter, and consider the motion of two point particles. As we will see, the resummation of the ladder series to all orders is possible in this frame thanks to the linear nature of the $\varphi T$ mixing.

\section{Ladder Resummation in Binary Systems}
\label{sec:2body}
To illustrate these effects more concretely, consider two point particles with masses $m_1$ and $m_2$ moving along worldlines $x_1 (\tau)$ and $x_2 (\tau)$. The relevant matter action is, 
\begin{align}
S_m  &= S_1 + S_2   \nonumber \\
S_A &=  \int d \tau_A \;\; \frac{m_A}{2} \left( - \frac{  \tilde{g}_{\mu \nu} (x_A) \,  u_A^\mu  u_A^\nu  }{ \tilde{e}_A} + \tilde{e}_A   \right)  
\end{align}
where $u_A ( \tau_A ) = \partial x_A (\tau_A) / \partial \tau_A$ is the particle four-velocity, $\tilde{g}_{\mu \nu} ( x_A ( \tau_A ))$ is the metric evaluated on the particle worldine, and $\tilde{e}_A (\tau_A )$ is an auxilliary field which can be integrated out straightforwardly,
\begin{equation}
\frac{\delta S_A}{\delta \tilde{e}_A} = 0 \;\;\;\; \Rightarrow \;\;\;\; \tilde{e}_A = \sqrt{- \tilde{g}_{\mu \nu} ( x_A) \,  u_A^\mu u_A^\nu} \, ,
\end{equation}
and is nothing more than the Jordan-frame norm of the four-velocity. We will continue to write $\tilde{e}_A$ explicitly because it makes the wordline reparametrization manifest\footnote{
In a general wordline gauge, $\tilde{e}_A ( \tau )$ depends on $\tau$, but we will only consider gauges in which $\tilde{e}_A$ is constant---this simplifies many subsequent expressions since $\partial_{\tau} \tilde{e}_A$ can be neglected. This restriction to constant $\tilde{e}_A$ gauges corresponds to only allowing rescalings $\tau \to \alpha \tau$, rather than full wordline diffeomorphisms, $\tau \to f ( \tau )$.   
} and will allow for a straightforward massless limit to be taken (see \eqref{eqn:massless}).  

The corresponding stress-energy of each particle is localised on their respective worldlines,
\begin{equation}
 T_A^{\mu \nu} (x) 
 = - m_A \int \frac{d \tau_A}{ \tilde e_A} \, u_A^\mu u_A^\nu \; \delta^{(4)} ( x - x_A ) \, . 
\end{equation}

The metric perturbation can be decomposed into $h_1^{\mu \nu} + h_2^{\mu \nu}$, so that \eqref{eqn:heomT} becomes a pair of equations,  
\begin{equation}
\Box  h_1^{\mu \nu} - \frac{1}{2} \eta^{\mu\nu} \Box h_1 = \frac{1}{M_P} T_1^{\mu \nu}  \;\; , \;\;\;\; \Box h_2^{\mu \nu} - \frac{1}{2} \eta^{\mu\nu} \Box h_2  = \frac{1}{M_P} T_2^{\mu \nu} \, ,
\label{eqn:heom2}
\end{equation}
and we can solve for the metric sourced around particle 1 and around particle 2 independently at this order.  

Exploiting the linearity of the $\varphi$ equation of motion \eqref{eqn:phieom}, we can similarly split the field $\varphi = \varphi_1 + \varphi_2$ and solve a pair of coupled equations,
\begin{align}
 \Box \varphi_1 &=  - \frac{ \con }{2 M_P} T_1 + \frac{ \dis }{\M^4} \nabla_\mu \left(  \nabla_\nu \varphi_2   T_1^{\mu \nu} \right) \, ,  \label{eqn:phi1fulleom} \\
 \Box \varphi_2 &=  - \frac{ \con }{2 M_P} T_2 + \frac{ \dis }{\M^4}  \nabla_\mu \left(  \nabla_\nu \varphi_1   T_2^{\mu \nu} \right) \, \, , \nonumber
\end{align}
where we have neglected any self-energy divergences\footnote{
Physically, these divergences would be regulated by (subleading) finite size effects.
} that arise from evaluating fields (and delta functions) directly on their own worldlines, i.e. terms of the form $\varphi_A T_A$. 
This should be read as particle A sourcing a field $\varphi_A (x)$, which can then also act as a background on which the other particle sources its field. In the absence of the disformal interaction, the equations decouple and the field around each particle is sourced independently. 

The goal is to solve the equations of motion \eqref{eqn:heom2} and \eqref{eqn:phi1fulleom} for $h^{\mu \nu} ( x )$ and $\phi (x)$ as functionals of the wordline trajectories, and then to solve the matter equations of motion,
\begin{equation}
 a_A^\mu + \tilde{\Gamma}^\mu_{\alpha \beta} ( x_A )\, u_A^{\alpha} u_A^{\beta} = 0  \, , 
\end{equation}
for $x_A ( \tau )$, where $\tilde{\Gamma}^\mu_{\alpha \beta} (x)$ are the Jordan-frame Christoffel symbols of $\tilde{g}_{\mu \nu} (x)$. We will separate $\tilde{\Gamma}^\mu_{\alpha \beta}$ into three pieces, namely the $\Gamma^{\mu}_{\alpha \beta}$ from the background metric (in our case $\eta_{\mu \nu}$), the linearised $\delta \Gamma^\mu_{\alpha \beta}$ from the Einstein-frame fluctuation $h_{\mu \nu}$, and the remaining scalar contributions. The geodesic equation is then,
\begin{equation}
  a_A^\mu + \Gamma^\mu_{\alpha \beta} ( x_A )\, u_A^{\alpha} u_A^{\beta} = F^\mu_{h \, A} + F^\mu_{\varphi \, A}  \, , 
  \label{eqn:geodesic}
\end{equation}
where $F^\mu_{h \, A} ( \tau ) = - \delta \Gamma^{\mu}_{\alpha \beta} ( x_A ) u_A^\alpha u_A^\beta $ is the usual gravitational force, and $F^\mu_{\varphi \, A} ( \tau ) = - ( \delta \tilde \Gamma^\mu_{\alpha \beta} (x_A) - \delta \Gamma^{\mu}_{\alpha \beta} (x_A) ) u_A^\alpha u_A^\beta$ is the scalar-mediated fifth force, given at leading order by,
\begin{align}
 F_{h \, A}^\mu ( \tau ) &= \frac{1}{M_P} \left( \frac{u_A^{\alpha} u_A^{\beta}}{2} \nabla^\mu h_{\alpha \beta} (x_A) -  u_A^{\beta} \partial_{\tau} h^{\mu}_{\beta} (x_A) \right)     \;\; ,  \label{eqn:Fh} \\
 F_{\varphi \, A}^\mu (\tau) &= - \frac{\con}{M_P} \left( \frac{ \tilde{e}_A^2}{2} \nabla^\mu \varphi (x_A) + u_A^{\mu} \partial_{\tau} \varphi (x_A) \right) \nonumber \\
 &\quad + \frac{\dis}{\M^4} \nabla^\mu \varphi (x_A) \left( \partial_{\tau}^2 \varphi ( x_A )  - a_A^\beta \nabla_\beta \varphi (x_A)  \right)  \, .  \label{eqn:Fp}
\end{align}
The gravitational and the conformal fifth forces are functionally very similar, however the disformal fifth force exhibits many novel characteristics. 
As we shall see, the disformal interaction leads to a memory effect which is effectively \emph{non-local} in time: if the matter distribution changes sufficiently quickly (e.g. if two particles have a sufficiently large relative velocity), then this memory effect is large, and any ``approximately static'' approximation will break down\footnote{
We believe this is one reason why the ladder screening mechanism presented here has not been observed before. The static approximation is usually a key ingredient when deriving screened profiles, e.g. Vainshtein, around single sources. 
}.   
In the remainder of this section we will solve the field equations for $h^{\mu \nu}$ and $\varphi$, and then in Section~\ref{sec:eg} compute these forces explicitly, and discuss the effects that they have on the trajectories of binary systems.

\subsection{Relativistic Motion}

It will prove useful to first define some notation which will allow us to massage \eqref{eqn:phi1fulleom} into a tractable form for arbitrary motion. Then in Section~\ref{sec:PN} we will focus on non-relativistic velocities and solve the equation in a Post-Newtonian expansion. 

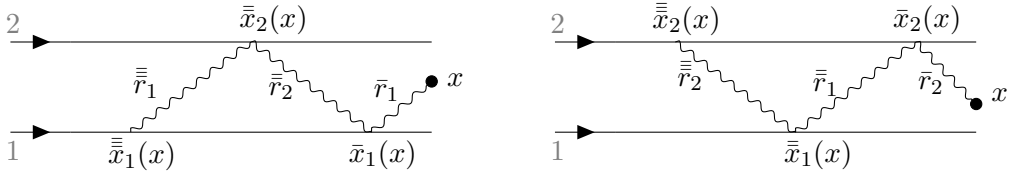
\begin{figure}
\begin{center}
		\begin{tikzpicture}[baseline=-0.6cm]
			\begin{feynman}
				\vertex (a1);
				\node [below=0.45cm of a1, dot] (v);
				\vertex [below=0.6cm of a1] (b1);
				\vertex [below=0.6cm of b1] (c1);
				\vertex [left=0.8cm of a1] (a2);
				\vertex [below=0.6cm of a2] (b2);
				\vertex [below=0.6cm of b2] (c2);
				\vertex [left=1.6cm of a2] (a3);
				\vertex [below=0.6cm of a3] (b3);
				\vertex [below=0.6cm of b3] (c3);
				\vertex [left=1.6cm of a3] (a4);
				\vertex [below=0.6cm of a4] (b4);
				\vertex [below=0.6cm of b4] (c4);
				\vertex [left=0.8cm of a4] (a5);
				\vertex [below=0.6cm of a5] (b5);
				\vertex [below=0.6cm of b5] (c5);
				\vertex [left=0.8cm of a5] (a6);
				\vertex [below=0.6cm of a6] (b6);
				\vertex [below=0.6cm of b6] (c6);

				\node at (+0.25, -0.5) {{ $x$ }};
				\node at (-0.65, -1.5) {{$ \bar{x}_1 (x)$}};
				\node at (-0.6, -0.65) {{$ \bar{r}_1$}};
				\node at (-2.1,  0.3) {{$ \bar{\bar{x}}_2 (x)$}};
				\node at (-2.0, -0.6) {{$ \bar{\bar{r}}_2$}};
				\node at (-3.85 ,  -1.5) {{$ \bar{\bar{\bar{x}}}_1 (x)$}};
				\node at (-3.8, -0.55) {{$ \bar{\bar{\bar{r}}}_1$}};

				\node at (-5.5, 0.25) {{\color{gray} $2$ }};
				\node at (-5.5, -1.45) {{\color{gray} $1$ }};
				
				\diagram*{
				(a6) -- [fermion] (a5) -- (a1),
				(c6) -- [fermion] (c5) -- (c1),
				(v) -- [photon] (c2) -- [photon] (a3) -- [photon] (c4),
				};
				
			\end{feynman}	
		\end{tikzpicture}
		\qquad
		\begin{tikzpicture}[baseline=-0.6cm]
			\begin{feynman}
				\vertex (a1);
				\node [below=0.75cm of a1, dot] (v);
				\vertex [below=0.6cm of a1] (b1);
				\vertex [below=0.6cm of b1] (c1);
				\vertex [left=0.8cm of a1] (a2);
				\vertex [below=0.6cm of a2] (b2);
				\vertex [below=0.6cm of b2] (c2);
				\vertex [left=1.6cm of a2] (a3);
				\vertex [below=0.6cm of a3] (b3);
				\vertex [below=0.6cm of b3] (c3);
				\vertex [left=1.6cm of a3] (a4);
				\vertex [below=0.6cm of a4] (b4);
				\vertex [below=0.6cm of b4] (c4);
				\vertex [left=0.8cm of a4] (a5);
				\vertex [below=0.6cm of a5] (b5);
				\vertex [below=0.6cm of b5] (c5);
				\vertex [left=0.8cm of a5] (a6);
				\vertex [below=0.6cm of a6] (b6);
				\vertex [below=0.6cm of b6] (c6);

				\node at (+0.25, -0.7) {{ $x$ }};
				\node at (-0.65, 0.3) {{$ \bar{x}_2 (x)$}};
				\node at (-0.6, -0.55) {{$ \bar{r}_2$}};
				\node at (-2.1,  -1.5) {{$ \bar{\bar{x}}_1 (x)$}};
				\node at (-2.0, -0.55) {{$ \bar{\bar{r}}_1$}};
				\node at (-3.85 , 0.3) {{$ \bar{\bar{\bar{x}}}_2 (x)$}};
				\node at (-3.8, -0.45) {{$ \bar{\bar{\bar{r}}}_2$}};

				\node at (-5.5, 0.25) {{\color{gray} $2$ }};
				\node at (-5.5, -1.45) {{\color{gray} $1$ }};
				
				\diagram*{
				(a6) -- [fermion] (a5) -- (a1),
				(c6) -- [fermion] (c5) -- (c1),
				(v) -- [photon] (a2) -- [photon] (c3) -- [photon] (a4),
				};
				
			\end{feynman}	
		\end{tikzpicture}
\end{center}
\caption{Iterated retarded positions on the wordlines. The exchanged signal could be a graviton or a scalar (or any massless fluctuation), and propagates along null geodesics between the two particles.}
\label{fig:notation}
\end{figure}

\paragraph{Kinematic Variables:}
At every $\tau$, each particle defines a privileged origin for our spacetime coordinates, in which $x_A = (0,0,0,0)$. We can describe this covariantly using the separation,
\begin{equation}
  r_A^\mu (x, \tau ) = x^\mu - x_A^\mu ( \tau ) \, .
  \label{eqn:retarded}
\end{equation}
Although the action is an integral over spacetime coordinates, $x^\mu$, and wordline time, $\tau$, thanks to the symmetries of the problem we know that solutions should only depend on the distances from the worldlines \eqref{eqn:retarded}.
Also, the causal structure of the background spacetime\footnote{
More generally, it is $g_{\mu \nu}$ which should be used to define the retarded $\rtau$, since it is the Einstein-frame metric which determines the causal cones for $\varphi$ and $h_{\mu\nu}$. 
} allows us to associate a unique $\tau$ on each worldline to every point $x^\mu$, namely the solution to,
\begin{equation}
 \eta_{\mu \nu} \, r_A^\mu ( x, \rtau_A ) r_A^\nu ( x, \rtau_A ) = 0 \, \;\; \Rightarrow \;\; \rtau_A (x) \, . 
\end{equation}
This gives a notion of the ``retarded positions'' and ``retarded separations'',
\begin{equation}
 \bar{x}_A^\mu (x) = x_A^{\mu} ( \rtau_A ) \, , \;\;\;\; \bar{r}_A^\mu (x) = r_A^\mu ( x, \rtau_A )  \, , 
\end{equation}
which depend on $x^\mu$ only. For a two body system, we can also define iterated retarded positions as follows:
\begin{equation}
 \bar{\bar{x}}_1^\mu (x) = \bar{x}_1^{\mu} ( \bar{x}_2 (x) )  \; ,\;\;\;\;  \bar{\bar{x}}_2^\mu (x) = \bar{x}_2^{\mu} ( \bar{x}_1 (x) )  \, . 
\end{equation}
Physically, $\bar{\bar{x}}_A (x)$ is the worldline position at which a null signal could be emitted from particle $A$ to the opposite particle, and then propagate from there to reach the spacetime point $x$. Similarly, the iterated retarded separations are,
\begin{align}
 \bar{\bar{r}}_1^\mu (x) &= \bar{r}_1^{\mu} ( \bar{x}_2 (x) )  \; \;\;\;\;  &\bar{\bar{r}}_2^\mu (x) &= \bar{r}_2^{\mu} ( \bar{x}_1 (x) )  \,  \nonumber   \\
 &=  \bar{x}_2 (x) - \bar{\bar{x}}_1 (x)   \; ,\;\;\;\;  &  &=  \bar{x}_1 (x) - \bar{\bar{x}}_2 (x)  \, .    
\end{align}
as depicted in Figure~\ref{fig:notation}.

Furthermore, we will use the particle 4-velocities to define a distance,
\begin{align}
 R_A (x, \tau) = \frac{ r_A \cdot u_A }{ \tilde{e}_A}  \, , 
 \label{eqn:RAdef}
\end{align}
and define the retarded spatial separations\footnote{
Note that since $\bar{r}_A^\mu$ is null by construction, we have the useful identity $\bar{R}_A (x) =  \sqrt{ \bar{r}_A^{\mu} \bar{P}_A^{\mu \nu} \bar{r}_A^{\nu} }$, where $P_A^{\mu\nu} (\tau) = \tilde{g}_{\mu\nu} + u_A^\mu u_A^\nu / \tilde{e}_A^2 $ is the projection tensor onto space-like components in the instantaneous rest frame of particle A at time $\tau$.
} $\bar{R}_A (x)$, $\bar{\bar{R}}_A (x)$, etc., as above. For inertial motion, $R_A$ does not depend on $\tau$ since it represents the spatial distance from the particle in the particle's own rest frame.  

As a final piece of notation, we will also replace $M_P$ and $\M^4$ with their corresponding constant length scales,
\begin{equation}
R_{S_A} = \frac{m_A}{8 \pi M_P^2}   \;\;\; , \;\;\; R_{V_A} = \left( \frac{m_A}{4 \pi \M^4} \right)^{1/3} \, ,
\end{equation}
which represent the Schwarzschild and Vainshtein radius of each mass.
With the notation set, we can now solve the equation of motions \eqref{eqn:heom2} and \eqref{eqn:phi1fulleom} fully relativistically.

\paragraph{Metric Fluctuations:}
Let us begin with the linearised Einstein equation \eqref{eqn:heom2}. The solution is well-known, and in terms of the retarded positions defined above reads, 
\begin{equation}
h^{\mu \nu}_A (x) = \frac{ R_{S_A} }{ \bar{R}_A (x) }  \; \frac{ u_A^\mu u_A^\nu - \frac{ e_A^2 }{2} g^{\mu \nu}}{ \tilde{e}_A^2 } \, ,
\label{eqn:eom_metric_fluc}
\end{equation}
(see Appendix~\ref{sec:ladder_explicit} for the derivation). The metric fluctuation falls off like $R_{S_A}/r$, where $r$ is the spatial separation between $x$ and the retarded position $\bar{x}_A (x)$. Note that the $e_A^2 = - \eta_{\mu\nu} u_A^\mu u_A^\nu$ in the numerator differs from $\tilde{e}_A$ only by terms which are $1/M_P$ or $1/\M^4$ suppressed, and thus contributes a correction only at next-to-leading order. 

\paragraph{Ladder Expansion:}
Now we turn to the scalar equation of motion \eqref{eqn:phi1fulleom}. 
Expanding the field as in \eqref{eqn:varphiexp}, one finds,
\begin{align}
\Box \varphi_1^{(0)} &= - \frac{ \con }{2M_P} T_1    \, ,  \label{eqn:ladder0} \\
\Box \varphi_1^{(1)} &=  \frac{ \dis }{ \M^4}  \nabla_\mu \left(   \nabla_\nu \varphi_2^{(0)}   T_1^{\mu \nu} \right)   \,  \label{eqn:ladder1}  ,  
\end{align}
and so on.    
These are straightforward to solve (we defer the algebra to Appendix~\ref{sec:ladder_explicit}), and the resulting contributions to the field profile are, 
\begin{align}
\frac{ \varphi_1^{(0)} ( x )}{M_P} &=  \, \frac{ - \con  R_{S_1} }{ \bar{R}_1 }   \, ,   
\,  \label{eqn:ladder0sol} \\
\frac{\varphi_1^{(1)} (x) }{M_P}  &= \frac{\dis \, R_{V_1}^3 }{ \bar{R}_1 } \frac{ \partial_{\rtau_1}^2 }{ \tilde{e}_1^2 } \left[  \frac{ - \con  R_{S_2} }{ \bar{\bar{R}}_2 }   \right]  \, .  \label{eqn:ladder1sol}
\end{align}
These are represented by the first two diagrams in Figure~\ref{fig:resum}. This procedure can be continued straightforwardly for $\varphi^{(2)}, \, \varphi^{(3)} \, , $ etc.
This is the relativstic generalisation of the results obtained in \cite{Brax:2018bow, Brax:2019tcy} in the Newtonian limit.

\paragraph{Ladder Resummation:}
Formally, the ladder series takes the form,
\begin{align}
\varphi_1 = \sum_{n=0}^\infty \varphi_1^{(n)}  \, , 
\end{align}
where the $n^{\rm th}$ term can be expressed recursively using the preceding terms,
\begin{equation}
\varphi_1^{(n)} =  \hat D_1 \left[ \varphi_2^{(n-1)} \right]  = \hat D_1 \hat D_2 \left[ \varphi_1^{(n-2)} \right] \, , 
\end{equation}
by using the two operators, 
\begin{equation}
\hat{D}_A \left[  f (x) \right]  =  \frac{ \dis \, R_{V_A}^3}{ \bar{R}_A (x) } \frac{ \partial_{\rtau_A}^2 f ( \bar{x}_A ) }{ \tilde e_A^2 } \,  \, . 
\end{equation}
This allows us to write,
\begin{align}
\varphi_1 (x) &= \sum_{n=0}^{\infty}  \left( \hat D_1 \hat D_2 \right)^n \left[  \varphi_1^{(0)} (x)  + \varphi_1^{(1)} (x)    \right]  \nn \\
&=  \frac{1}{1- \hat D_1 \hat D_2} \left[  \varphi_1^{(0)} (x) + \varphi_1^{(1)} (x)     \right] \, .   \label{eqn:ladderresummed}
\end{align}
This final expression is purely formal, but represents how the infinite series of diagrams should be resummed (shown graphically in Figure~\ref{fig:resum}). 

The full field profile, with no ladder expansion, therefore obeys the differential equation:
\begin{equation}
\varphi_1 -  \hat D_1 \hat D_2 \left[ \varphi_1 \right]  = \varphi_1^{(0)} + \varphi_1^{(1)} \, .
\label{eqn:eomresum}
\end{equation} 
This is the key result for the two-body problem. It allows us to take the first terms in the ladder expansion, and use them as a \emph{source} with which to compute the full (non-perturbative) result. 
 
To make this explicit, consider the ansatz, 
\begin{equation}
\frac{ \varphi_A (x) }{M_P} = - \frac{ \con R_{S_A} }{\bar{R}_A (x)}  G_A ( \rtau_A (x) )
\label{eqn:phiansatzRela}
\end{equation}
so that \eqref{eqn:eomresum} becomes, 
\begin{equation}
 G_1 (  \rtau_1  ) -  \dis^2 \, \nu  \frac{ \partial_{\rtau_1}^2 }{ \tilde{e}_1^2 } \left[    \frac{ R_{V_1}^3 }{ \bar{R}_2 ( \bar{x}_1 ) } \frac{ \partial_{ \bar{\bar{\tau}}_2}^2 }{ \tilde{e}_2^2 } \left[ \frac{ R_{V_1}^3}{  \bar{\bar{R}}_1 ( \bar{x}_1 ) } G_1 (  \bar{\bar{ \bar{ \tau }}}_1  )   \right] \right] =  1  + \dis \, \nu  \frac{\partial^2_{\rtau_1}}{\tilde{e}_1^2} \left[ \frac{ R_{V_1}^3 }{ \bar{R}_2 ( \bar{x}_1 )}  \right]
 \label{eqn:GeqnRela}
\end{equation}
where $\nu = m_2 / m_1$ is the mass ratio. This equation determines the effective Newton's constant for the scalar force as a functional of the worldline trajectories $x_A (\tau)$ only. Note that for small disformal interactions ($d \to 0$), or in the one-body limit\footnote{
Note that the one-body limit is $\nu \to 0$ with $\tilde{e}_A$ held fixed, while the massless limit is $\nu \to 0$ with $m_2/\tilde{e}_2$ held fixed---we will return to this at the end of this section.  
} ($\nu \to 0$), this returns the standard result for a conformally coupled scalar field, $G_1 = 1$. When including disformal effects in a relativistic two-body system, $G_1$ is instead given by the full equation \eqref{eqn:GeqnRela}.     

Crucially, note that \eqref{eqn:GeqnRela} is a \emph{functional} differential equation, because the operator $\hat{D}_A$ returns the function evaluated at a \emph{different} argument---in this case it maps $\bar{x}_A$ to the $\bar{\bar{\bar{x}}}_A$ shown in Figure~\ref{fig:notation}. This lies at the heart of what complicates the two-body problem: the interaction is effectively non-local in time, since it couples the current position/velocity of the particle to the position/velocity of the particle in its retarded past---the system has some memory of its past configuration.  
 
One might object at this stage that we have unfairly inverted a differential operator, and therefore have introduced an unnecessary ambiguity in our $G_1$ solution. That is to say, any solution to,
\begin{equation}
f (x) - \hat D_1  \hat D_2 [ f (x) ] = 0 
\label{eqn:homo}
\end{equation}
can now be added to $\varphi_1 (x)$ and \eqref{eqn:eomresum} will still be satisfied. 
Since $1 - \hat D_1  \hat D_2$ is a functional operator\footnote{
Due to the functional nature of the equation, it is not clear how many independent solutions \eqref{eqn:homo} will have. 
In some cases, there are four independent Airy-like solutions, found by solving, 
\begin{equation}
\frac{ \partial_{\rtau_A}^2 }{ \tilde{e}_A^2 }  f \left(  \bar{x}_A \right)  = + \frac{ \bar{R}_A (x)}{ \dis \, R_{V_A}^3 }  \, f \left( x \right)  \; \;\;\; \text{or} \;\;\;\; \frac{ \partial_{\rtau_A}^2 }{ \tilde{e}_A^2 }  f \left(  \bar{x}_A \right)  = - \frac{ \bar{R}_A (x)}{ \dis \, R_{V_A}^3 }  \, f \left( x \right)  \; .
\end{equation}
}, it is not immediately clear whether it has a non-trivial kernel (i.e. whether there are actually any solutions to \eqref{eqn:homo}), however there are three possible cases:
\begin{itemize}

\item The trajectories $x_A ( \tau )$ are such that \eqref{eqn:homo} has no solutions, in which case \eqref{eqn:GeqnRela} gives a unique solution for $G_1$.  

\item Alternatively, \eqref{eqn:homo} might have solutions, but they do not obey the retarded boundary conditions we have imposed on $\varphi$ (in going from (\ref{eqn:ladder0}, \ref{eqn:ladder1}) to (\ref{eqn:ladder0sol}, \ref{eqn:ladder1sol})). Such solutions are inconsistent with the ladder expansion and therefore should be discarded, in which case \eqref{eqn:GeqnRela} gives a unique solution for $G_1$. 

\item Finally, \eqref{eqn:homo} could have solutions which \emph{do} obey the asymptoptic boundary conditions, and \eqref{eqn:GeqnRela} therefore only provides $G_1$ up to undetermined constants of integration. These undetermined constants are unavoidable, and correspond to an ambiguity in the Borel resummation of the ladder series (due to the so-called \emph{Stokes phenomenon}). Physically, they represent a non-perturbative correction to the ladder series (e.g. a contribution like $e^{-r^2}$), which is naively invisible to perturbation theory (e.g. $e^{-r^2} = 0 + 0 r^{-2} + 0 r^{-4} + ... $ at large $r$), but which arises from our resummed equation via resurgence. The fact that our perturbative resummation can access this non-perturbative information is remarkable, and can be traced to instanton-like contributions from other saddle points in the effective action. We will discuss these features further in Section~\ref{sec:head-on}, with the aid of a fully worked example. 

\end{itemize}

\begin{figure}
\begin{flushleft}
\qquad  $\; \varphi_1 (x) \;\; = \;\;\;\;$
		  \begin{tikzpicture}[baseline=-0.6cm]
			\begin{feynman}
				\vertex (a1);
				\node [below=0.45cm of a1, dot] (v);
				\vertex [below=0.5cm of a1] (b1);
				\vertex [below=0.5cm of b1] (c1);
				\vertex [left=0.4cm of a1] (a2);
				\vertex [below=0.5cm of a2] (b2);
				\vertex [below=0.5cm of b2] (c2);
				\vertex [left=0.8cm of a2] (a3);
				\vertex [below=0.5cm of a3] (b3);
				\vertex [below=0.5cm of b3] (c3);
				\node at (-0.5, -1.25) {$c$};
				\node at (-1.2, 0.25) {{\color{gray} $2$ }};
				\node at (-1.2, -1.25) {{\color{gray} $1$ }};
				
				\diagram*{
				(a3) -- [fermion] (a2) -- (a1),
				(c3) -- [fermion] (c2) -- (c1),
				(v) -- [scalar] (c2),
				};
				
			\end{feynman}		
		\end{tikzpicture}
		+		
		\begin{tikzpicture}[baseline=-0.6cm]
			\begin{feynman}
				\vertex (a1);
				\node [below=0.45cm of a1, dot] (v);
				\vertex [below=0.5cm of a1] (b1);
				\vertex [below=0.5cm of b1] (c1);
				\vertex [left=0.4cm of a1] (a2);
				\vertex [below=0.5cm of a2] (b2);
				\vertex [below=0.5cm of b2] (c2);
				\vertex [left=0.8cm of a2] (a3);
				\vertex [below=0.5cm of a3] (b3);
				\vertex [below=0.5cm of b3] (c3);
				\vertex [left=0.8cm of a3] (a4);
				\vertex [below=0.5cm of a4] (b4);
				\vertex [below=0.5cm of b4] (c4);
				\node at (-1.25,  0.25) {$c$};
				\node at (-0.5, -1.25) {$d$};
				\node at (-2.0, 0.25) {{\color{gray} $2$}};
				\node at (-2.0, -1.25) {{\color{gray} $1$}};

				\diagram*{
				(a4) -- [fermion] (a3) -- (a1),
				(c4) -- [fermion] (c3) -- (c1),
				(v) -- [scalar] (c2) -- [scalar] (a3),
				};				
			\end{feynman}		
		\end{tikzpicture}
		+
		\begin{tikzpicture}[baseline=-0.6cm]
			\begin{feynman}
				\vertex (a1);
				\node [below=0.45cm of a1, dot] (v);
				\vertex [below=0.5cm of a1] (b1);
				\vertex [below=0.5cm of b1] (c1);
				\vertex [left=0.4cm of a1] (a2);
				\vertex [below=0.5cm of a2] (b2);
				\vertex [below=0.5cm of b2] (c2);
				\vertex [left=0.8cm of a2] (a3);
				\vertex [below=0.5cm of a3] (b3);
				\vertex [below=0.5cm of b3] (c3);
				\vertex [left=0.8cm of a3] (a4);
				\vertex [below=0.5cm of a4] (b4);
				\vertex [below=0.5cm of b4] (c4);
				\vertex [left=0.8cm of a4] (a5);
				\vertex [below=0.5cm of a5] (b5);
				\vertex [below=0.5cm of b5] (c5);
				\node at (-1.25,  0.25) {$d$};
				\node at (-0.5, -1.25) {$d$};
				\node at (-2.0, -1.25) {$c$};
				\node at (-2.8, 0.25) {{\color{gray} $2$ }};
				\node at (-2.8, -1.25) {{\color{gray} $1$ }};
				
				\diagram*{
				(a5) -- [fermion] (a4) -- (a1),
				(c5) -- [fermion] (c4) -- (c1),
				(v) -- [scalar] (c2) -- [scalar] (a3) -- [scalar] (c4),
				};				
			\end{feynman}		
		\end{tikzpicture}
		$\; + \; ...$  \\[10pt]
\qquad  \phantom{$\; \varphi_1 (x) \;\;$}				
		$= \; \displaystyle \sum_{2n} \;$ 
		\begin{tikzpicture}[baseline=-0.6cm]
			\begin{feynman}
				\vertex (a1);
				\node [below=0.45cm of a1, dot] (v);
				\vertex [below=0.5cm of a1] (b1);
				\vertex [below=0.5cm of b1] (c1);
				\vertex [left=0.4cm of a1] (a2);
				\vertex [below=0.5cm of a2] (b2);
				\vertex [below=0.5cm of b2] (c2);
				\vertex [left=0.4cm of a2] (a3);
				\vertex [below=0.5cm of a3] (b3);
				\vertex [below=0.5cm of b3] (c3);
				\vertex [left=0.8cm of a3] (a4);
				\vertex [below=0.5cm of a4] (b4);
				\vertex [below=0.5cm of b4] (c4);
				\vertex [left=0.4cm of a4] (a5);
				\vertex [below=0.5cm of a5] (b5);
				\vertex [below=0.5cm of b5] (c5);
				\vertex [left=0.8cm of a5] (a6);
				\vertex [below=0.5cm of a6] (b6);
				\vertex [below=0.5cm of b6] (c6);
				\vertex [left=0.8cm of a6] (a7);
				\vertex [below=0.5cm of a7] (b7);
				\vertex [below=0.5cm of b7] (c7);
				\node at (-1.25, -0.5) {...};
				\node at (-2.0,  0.25) {$d$};
				\node at (-0.5, -1.25) {$d$};
				\node at (-2.8, -1.25) {$c$};
				\node at (-3.6, 0.25) {{\color{gray} $2$ }};
				\node at (-3.6, -1.25) {{\color{gray} $1$ }};
				
				\diagram*{
				(a7) -- [fermion] (a6) -- (a1),
				(c7) -- [fermion] (c6) -- (c1),
				(v) -- [scalar] (c2) -- [scalar] (b3),
				(c6) -- [scalar] (a5) -- [scalar] (b4),
				};				
			\end{feynman}		
		\end{tikzpicture}
		$\; +\displaystyle \sum_{2n+1} \;$
		\begin{tikzpicture}[baseline=-0.6cm]
			\begin{feynman}
				\vertex (a1);
				\node [below=0.45cm of a1, dot] (v);
				\vertex [below=0.5cm of a1] (b1);
				\vertex [below=0.5cm of b1] (c1);
				\vertex [left=0.4cm of a1] (a2);
				\vertex [below=0.5cm of a2] (b2);
				\vertex [below=0.5cm of b2] (c2);
				\vertex [left=0.4cm of a2] (a3);
				\vertex [below=0.5cm of a3] (b3);
				\vertex [below=0.5cm of b3] (c3);
				\vertex [left=0.8cm of a3] (a4);
				\vertex [below=0.5cm of a4] (b4);
				\vertex [below=0.5cm of b4] (c4);
				\vertex [left=0.4cm of a4] (a5);
				\vertex [below=0.5cm of a5] (b5);
				\vertex [below=0.5cm of b5] (c5);
				\vertex [left=0.8cm of a5] (a6);
				\vertex [below=0.5cm of a6] (b6);
				\vertex [below=0.5cm of b6] (c6);
				\vertex [left=0.8cm of a6] (a7);
				\vertex [below=0.5cm of a7] (b7);
				\vertex [below=0.5cm of b7] (c7);
				\node at (-1.25, -0.5) {...};
				\node at (-2.0, -1.25) {$d$};
				\node at (-0.5, -1.25) {$d$};
				\node at (-2.8, 0.25) {$c$};
				\node at (-3.6, 0.25) {{\color{gray} $2$ }};
				\node at (-3.6, -1.25) {{\color{gray} $1$ }};

				\diagram*{
				(a7) -- [fermion] (a6) -- (a1),
				(c7) -- [fermion] (c6) -- (c1),
				(v) -- [scalar] (c2) -- [scalar] (b3),
				(a6) -- [scalar] (c5) -- [scalar] (b4),
				};				
			\end{feynman}		
		\end{tikzpicture}
		\\[5pt]
\qquad \phantom{$\; \varphi_1 (x) \;\;$}				
		$= \;\;$ 
		\begin{tikzpicture}[baseline=-1.2cm]
			\begin{feynman}
				\vertex (a1);
				\vertex [below=0.6cm of a1] (b1);
				\vertex [below=0.5cm of b1] (c1);
				\vertex [below=0.5cm of c1] (d1);				
				\vertex [below=0.6cm of d1] (e1);				
				\vertex [left=0.8cm of a1] (a2);
				\vertex [below=0.6cm of a2] (b2);
				\vertex [below=0.5cm of b2] (c2);
				\vertex [below=0.5cm of c2] (d2);				
				\vertex [below=0.6cm of d2] (e2);				
				\vertex [left=0.8cm of a2] (a3);
				\vertex [below=0.6cm of a3] (b3);
				\vertex [below=0.5cm of b3] (c3);
				\vertex [below=0.5cm of c3] (d3);				
				\vertex [below=0.6cm of d3] (e3);			
				\vertex [left=0.8cm of a3] (a4);
				\vertex [below=0.6cm of a4] (b4);
				\vertex [below=0.5cm of b4] (c4);
				\vertex [below=0.5cm of c4] (d4);				
				\vertex [below=0.6cm of d4] (e4);		
				\vertex [left=0.8cm of a4] (a5);
				\vertex [below=0.6cm of a5] (b5);
				\vertex [below=0.5cm of b5] (c5);
				\vertex [below=0.5cm of c5] (d5);	
				\vertex [below=0.6cm of d5] (e5);			
				\vertex [left=0.8cm of a5] (a6);
				\vertex [below=0.6cm of a6] (b6);
				\vertex [below=0.5cm of b6] (c6);
				\vertex [below=0.5cm of c6] (d6);				
				\vertex [below=0.6cm of d6] (e6);			
				\vertex [left=0.8cm of a6] (a7);
				\vertex [below=0.6cm of a7] (b7);
				\vertex [below=0.5cm of b7] (c7);
				\vertex [below=0.5cm of c7] (d7);				
				\vertex [below=0.6cm of d7] (e7);			
				\vertex [left=0.8cm of a7] (a8);
				\vertex [below=0.6cm of a8] (b8);
				\vertex [below=0.5cm of b8] (c8);
				\vertex [below=0.5cm of c8] (d8);				
				\vertex [below=0.6cm of d8] (e8);			
				\vertex [left=0.8cm of a8] (a9);
				\vertex [below=0.6cm of a9] (b9);
				\vertex [below=0.5cm of b9] (c9);
				\vertex [below=0.5cm of c9] (d9);				
				\vertex [below=0.6cm of d9] (e9);			
				\vertex [below=0.25cm of d5] (d5l);
				\node [below=0.2cm of d2, dot] (v3);
 				\vertex [above right=0.2cm of b5] (b5u);
 				\vertex [above right=0.2cm of e7] (e7u);
  				\vertex [below =0.3cm of d4] (e4u);
 				\node [right=0.4cm of e4u, dot] (v4);
				\vertex [right=0.4cm of e3] (e3r);
				
				\node at (d5l) {$-$};
				\node at (-3.8, 0.25) {{\color{gray} $2$ }};
				\node at (-3.8, -0.85) {{\color{gray} $1$ }};
				\node at (-5.4, -1.35) {{\color{gray} $2$ }};
				\node at (-5.4, -2.45) {{\color{gray} $1$ }};
				\node at (-2.2, -1.35) {{\color{gray} $2$ }};
				\node at (-2.2, -2.45) {{\color{gray} $1$ }};
				\node at (-1.6, -1.35) { $d$ };
				\node at (-1.2, -2.45) { $d$ };
				
				\diagram*{
				(c1) -- (c9),
				(a6) -- [fermion] (a5) -- (a4),
				(b6) -- [fermion] (b5) -- (b4),
				(d4) -- [fermion] (d3) -- (d2),
				(e4) -- [fermion] (e3) -- (e2),
				(v3) -- [scalar] (e3r) -- [scalar] (d3) -- [scalar] (v4),
				(d8) -- [fermion] (d7) -- (d6),
				(e8) -- [fermion] (e7) -- (e6),
				};				
			\end{feynman}		
		\end{tikzpicture}
		$\;\;\; \displaystyle \Bigg( \;$
				\begin{tikzpicture}[baseline=-0.6cm]
			\begin{feynman}
				\vertex (a1);
 				\node [below=0.45cm of a1, dot] (v);
				\vertex [below=0.5cm of a1] (b1);
				\vertex [below=0.5cm of b1] (c1);
				\vertex [left=0.4cm of a1] (a2);
				\vertex [below=0.5cm of a2] (b2);
				\vertex [below=0.5cm of b2] (c2);
				\vertex [left=0.8cm of a2] (a3);
				\vertex [below=0.5cm of a3] (b3);
				\vertex [below=0.5cm of b3] (c3);
				\node at (-0.5, -1.25) {$c$};
				\node at (-1.2, 0.25) {{\color{gray} $2$ }};
				\node at (-1.2, -1.25) {{\color{gray} $1$ }};
				
				\diagram*{
				(a3) -- [fermion] (a2) -- (a1),
				(c3) -- [fermion] (c2) -- (c1),
				(v) -- [scalar] (c2),
				};
				
			\end{feynman}		
		\end{tikzpicture}
		$\;\; + \;$		
		\begin{tikzpicture}[baseline=-0.6cm]
			\begin{feynman}
				\vertex (a1);
 				\node [below=0.45cm of a1, dot] (v);
				\vertex [below=0.5cm of a1] (b1);
				\vertex [below=0.5cm of b1] (c1);
				\vertex [left=0.4cm of a1] (a2);
				\vertex [below=0.5cm of a2] (b2);
				\vertex [below=0.5cm of b2] (c2);
				\vertex [left=0.8cm of a2] (a3);
				\vertex [below=0.5cm of a3] (b3);
				\vertex [below=0.5cm of b3] (c3);
				\vertex [left=0.8cm of a3] (a4);
				\vertex [below=0.5cm of a4] (b4);
				\vertex [below=0.5cm of b4] (c4);
				\node at (-1.25,  0.25) {$c$};
				\node at (-0.5, -1.25) {$d$};
				\node at (-2.0, 0.25) {{\color{gray} $2$}};
				\node at (-2.0, -1.25) {{\color{gray} $1$}};

				\diagram*{
				(a4) -- [fermion] (a3) -- (a1),
				(c4) -- [fermion] (c3) -- (c1),
				(v) -- [scalar] (c2) -- [scalar] (a3),
				};				
			\end{feynman}		
		\end{tikzpicture}
		$ \displaystyle \Bigg) \;\;$ 
\end{flushleft}
\caption{Resummation of the ladder expansion shown diagrammatically. The first two diagrams on the top line correspond to $\varphi_1^{(0)}$ and $\varphi_1^{(1)}$ (given in \eqref{eqn:ladder0} and \eqref{eqn:ladder1}), namely the field sourced by particle 1 at leading and next-to-leading order in the ladder expansion. The two sums on the second line correspond to the two terms in \eqref{eqn:ladderresummed}, and denote an even/odd number of disformal insertions respectively. The fraction on the third line corresponds to the $1/(1-\hat{D}_1 \hat{D}_2)$ resummation.}
\label{fig:resum}
\end{figure}

\paragraph{Different regimes:}
The equation \eqref{eqn:GeqnRela} is very difficult to solve in general due to its functional nature. 
However, there are two limiting cases in which we can make some analytic progress:
\begin{itemize}

 \item[(i)] In the ladder expansion regime, $r^3 \gg v^2 R_V^3$, then the disformal terms in \eqref{eqn:GeqnRela} are small and can be treated perturbatively, leading to,
 \begin{equation}
  G_1^{(0)} ( \bar{\tau}_1 ) = 1  \, , \;\;\;\; G_1^{(1)} ( \bar{\tau}_1 ) =  \dis \, \nu  \frac{\partial^2_{\rtau_1}}{ \tilde{e}_1^2} \left[ \frac{ R_{V_1}^3 }{ \bar{R}_2 ( \bar{x}_1 )}  \right]  \, , \;\;\; \text{etc.} 
  \label{eqn:Gladder}
 \end{equation}
 which reproduces the ladder expansion $\varphi^{(0)} + \varphi^{(1)} + ...$ derived above. 

 \item[(ii)] Deep inside the ladder resummation regime, $r^3 \ll v^2 R_V^3$, the disformal terms in \eqref{eqn:GeqnRela} dominate, and we can treat the first $G(\bar{x}_1)$ term as a small perturbation. This reduces the functional differential equation to an ordinary fourth-order differential equation, general solutions to which can be parametrised in terms of four undetermined constants. One solution is given by,
\begin{equation}
 G_1 (  \bar{\tau}_1 ( x_2 ( \tau )  ) )   = - \frac{ \bar{R}_1 ( x_2 ( \tau ) ) }{\dis \, R_{V_1}^3 } \, \frac{ \tau^2 }{ \tilde{e}_2^2} \,  , 
 \label{eqn:GPIeg}
\end{equation} 
 and so the scalar field experienced by particle 2 due to particle 1 is, 
 \begin{equation}
  \frac{ \varphi_1 ( x_2 ( \tau ) ) }{M_P} = - \frac{ \con R_{S_1} }{ \dis \, R_{V_1}^3 } \, \frac{ \tau^2 }{ \tilde{e}_2^2 } \, \sim \frac{\con}{\dis} \, H_0^2 \tau^2  
 \end{equation}
plus up to four undetermined constants (which should be matched to the ladder expansion). This already suggests that, if the scalar field is initially small, then it will grow very slowly (on Hubble timescales if $\phi$ is dark energy) as a result of the two-body dynamics, resulting in vastly suppressed fifth forces.
   
\end{itemize}

From the point of view of solar system tests, two of the most interesting setups are the deflection of a massless particle past a fixed heavy object (such as photons past the Sun), and the non-relativistic orbits of light masses around a heavy host (i.e. planetary orbits). We will comment briefly on the massless limit below, but the remainder of this work will be focussed primarily on how the Newtonian orbits of binary systems are affected by the presence of a disformally coupled scalar field, solving \eqref{eqn:GeqnRela} within a Post-Newtonian expansion.

\paragraph{Massless Particles:}
The massless limit is taken by sending,
\begin{equation}
 m_A \to 0  \;\;\;\; \text{with} \;\;\;\; \frac{ m_A }{ \tilde{e}_A } = \tilde{E}_A \;\; \text{fixed} \; . 
 \label{eqn:massless}
\end{equation}
The stress-energy tensor for a massless particle is then,
\begin{equation}
 T_A^{\mu \nu} \to -  \int d \tau_A \, \tilde{E}_A \, u_A^\mu u_A^\nu \, \delta^{(4)} ( x - x_A ( \tau_A ) )
\end{equation}
where $\tilde{E}_A$ inherits the transformation properties of $\tilde{e}_A$ to maintain wordline reparametrization invariance, and $u_A^\mu$ is now null ($\tilde{g}_{\mu \nu} u_A^\mu u_A^\nu = 0$) on integrating out $\tilde{E}_A$. The Einstein-frame trace of the stress-energy then vanishes at leading order, 
\begin{equation}
 T_A = 0 + \mathcal{O} \left( \frac{R_S}{r}   \right)
\end{equation}
where the subleading corrections come from the difference between $\tilde{g}_{\mu \nu}$ and $g_{\mu \nu}$ (i.e. $u_A^\mu$ is not null in the Einstein-frame, $g_{\mu \nu} u_A^\mu u_A^\nu \neq 0$). 

Let us suppose that particle 2 is massless. At leading order $\varphi_2^{(0)}$ then vanishes, and \eqref{eqn:GeqnRela} becomes, 
\begin{equation}
  G_1 ( \bar{\tau}_1 ) - 
  \dis^2 \, \frac{ \partial_{\rtau_1}^2 }{ \tilde{e}_1^2 } \left[    \frac{ R_{V_2}^3 }{ ( \bar{x}_1 - \bar{\bar{x}}_2 ) \cdot u_2 }  \partial_{ \bar{\bar{\tau}}_2}^2  \left[ \frac{ R_{V_1}^3}{  \bar{\bar{R}}_1 ( \bar{x}_1 ) } G_1 (  \bar{\bar{\bar{\tau}}}_1 ) )   \right] \right] 
  =  1  \,  .
  \label{eqn:GeqnRelaMassless}
\end{equation}
where the Vainshtein radius for the massless particle is now determined by its energy $R_{V_2}^3 = \tilde{E}_2/\M^4$.
Note that it is therefore \emph{not} the case that a massless particle will be somehow immune to the disformal coupling---it will be affected just as any massive particle would, and there is no violation of the equivalence principle. 

Let us further work in the rest frame of particle 1, and suppose that the impact parameter is orthogonal to the velocity of particle 2,
\begin{align}
 x_1^\mu = \left( \begin{array}{c}
               \tau_1 \\
               0 \\
               0 \\
               0
                  \end{array}
   \right)  \;\;, \;\;\;\; x_2^{\mu} = \left(  \begin{array}{c}
                                                \tau_2  \\
                                                b  \\
                                                0  \\
                                                \tau_2 
                                               \end{array}
     \right)
\end{align}
There is then a particularly simple relation between $\rtau_1$ and $\bar{\bar{\bar{\tau_1}}}$, 
\begin{equation}
\rtau_1 \, \bar{\bar{\bar{\tau}}}_1 = - b^2
\end{equation}
which suggests that it might be possible to solve the functional equation \eqref{eqn:GeqnRelaMassless} exactly. 
Pursuing this direction further would allow connection with the \emph{Cassini} measurements of the Shapiro time delay \cite{Bertotti:2003rm}, and provide an independent constraint on the $(\con, \, \dis)$ parameter space.
We leave this for future work, and turn now to the problem of two non-relativistic bodies.

\subsection{Post-Newtonian Expansion}
\label{sec:PN}

From here on, we will consider the scalar field sourced by non-relativistic particles. Not only is this case phenomenologically relevant, but it achieves an important technical simplification: it remedies the functional nature of \eqref{eqn:GeqnRela}. 
In particular, in the Newtonian limit the difference between the retarded position $\bar{\mathbf{x}}_A (t, \mathbf{x} ) $ and the coordinate time position $ \mathbf{x}_A (t) $ is small, and so equation~\eqref{eqn:GeqnRela} becomes an \emph{ordinary} differential equation for $G_1 ( t )$ (at leading order in $v_A^2$). 

\paragraph{Newtonian Limit:}
Firstly, we fix the worldline gauges $\tilde e_1 = \tilde e_2 = 1$, so that,
\begin{equation}
 u_A^0 ( \tau ) = 1 + \mathcal{O} \left( v_A^2 , h_{\mu \nu}  \right) \, , 
\end{equation}
where $\mathbf{v}_A$ is the spatial part of $u_A^\mu$, which we now treat as a small parameter\footnote{
Recall that we use units in which the speed of light is unity, and so in this gauge $v_A = | \mathbf{v}_A |$ represents the speed of the particle relative to light. 
}. This allows us to write,  
\begin{equation}
 \bar{\bar{R}}_2 = | \bar{\mathbf{x}}_1  - \bar{\bar{\mathbf{x}}}_2   |  + \mathcal{O} \left( v_A^2 , h_{\mu \nu}  \right) \, , \;\;\;\;  \bar{\bar{\bar{R}}}_1 = | \bar{\bar{\mathbf{x}}}_2  - \bar{\bar{\bar{\mathbf{x}}}}_1   |  + \mathcal{O} \left( v_A^2 , h_{\mu \nu}  \right) \, . 
\end{equation}
The retarded times can also be expanded,
\begin{equation}
 \bar{\tau}_A (t , \mathbf{x} )  = t -  | \mathbf{x} - \mathbf{x}_A ( t ) |  + \mathcal{O} ( v_A^2 )  \, .
\end{equation}
and so the retarded time derivatives can be written as coordinate-time derivatives, 
\begin{equation}
\frac{\partial}{\partial \bar \tau_A} = \frac{\partial}{\partial t}  + \mathcal{O} ( v_A ) 
\end{equation}
and the retarded positions can be written as coordinate-time postitions,
\begin{equation}
 \bar{x}_A^0 (t, \mathbf{x} ) = t - | \mathbf{x} - \mathbf{x}_A ( t ) | + \mathcal{O} (v_A^2) \; , \;\;\;\; \bar{\mathbf{x}}_A (t, \mathbf{x} ) = \mathbf{x}_A (t) + \mathcal{O} ( \mathbf{v}_A)
\end{equation}
where we have assumed all spatial separations are such that $| \mathbf{x} - \mathbf{x}_A (t) | \partial_t  \ll 1 $.

\paragraph{Metric Fluctuations:}
In this limit, the metric perturbations are now given by the usual Newtonian potential,
\begin{equation}
 h^{00}_A (t, \mathbf{x}) =  -  \frac{ 2 R_{S_A} }{ | \mathbf{x} - \mathbf{x}_A (t) |  }  \, , \;\;\;\;  h^{ij}_A (t, \mathbf{x}) =  -  \frac{ 2 R_{S_A} }{ | \mathbf{x} - \mathbf{x}_A (t) |  } \; \delta^{ij}  \,  ,  \;\;\;\;  h^{0i}_A ( t , \mathbf{x} ) = 0 \, . 
\end{equation}
Since we do not go beyond $\mathcal{O} \left( R_S / r   \right)$, it is sufficient to truncate $h^{\mu\nu}$ at this order. To capture the first GR corrections to Newtonian orbits, one must include the $\mathcal{O} (v^2)$ corrections here.

\paragraph{Scalar Fluctuations:}
The scalar field profile \eqref{eqn:phiansatzRela} in the Newtonian limit takes the form\footnote{
Similarly, next-to-leading-order corrections from the scalar would require keeping $\mathcal{O} ( v_A^2)$ terms in \eqref{eqn:phiansatzPN}. 
},
\begin{equation}
 \varphi_A (t, \mathbf{x} ) = - \frac{\con R_{S_A} }{ | \mathbf{x} - \mathbf{x}_A (t) | } \, G_A (t)
\label{eqn:phiansatzPN}
\end{equation}
where the resummed ladder equation \eqref{eqn:GeqnRela} determines $G_1 (t)$,
\begin{equation}
G_1 (t) - \dis^2  \nu \partial_t^2 \left( \frac{R_{V_1}^3}{| \mathbf{x}_{12} (t) |} \partial_t^2 \left(  \frac{R_{V_1}^3 }{| \mathbf{x}_{12} (t) |}  G_1 (t) \right)     \right)     = 1  +  \dis \nu  \partial_t^2 \left[  \frac{R_{V_1}^3}{ | \mathbf{x}_{12} ( t ) |  }  \right] \, ,
\label{eqn:Geqn}
\end{equation}
at leading order in the velocities, where $\mathbf{x}_{12} = \mathbf{x}_{1} - \mathbf{x}_2$ is the relative spatial separation of the particles, and we can now treat both terms in $G_1$ as evaluated at the same coordinate time $t$. Note that we have discarded time derivatives of $G_1 (t)$ as velocity-suppressed, but retained the leading disformal contributions (which $\sim v^2 R_V^3 / r^3$, since the small $v$ can be compensated by a large $R_V/r$).  

The Post-Newtonian expansion has successfully reduced the functional differential equation \eqref{eqn:GeqnRela} to the ordinary differential equation \eqref{eqn:Geqn}, which is now amenable to a wider range of analytic and computational techniques. However, it is still a fourth order equation with functional coefficients, and so finding a closed expression for $G_1 (t)$ in terms of a general $\mathbf{x}_{12} (t)$ remains challenging. In general, we expect that a solution can be parametrised in terms of four constants of integration,
\begin{equation}
  G_1 (t) = G^{\rm PI}_1 (t) + \sum_n \, c_n \, \frac{ | \mathbf{x}_{12} | }{R_{S_1}}  \; f_n ( t )   \, , 
  \label{eqn:Gsol}
\end{equation}
where $f_n$ are the four independent solutions to the generalized Airy equations,
\begin{equation}
 \partial_t^2 f (t) = + \frac{ | \mathbf{x}_{12} (t) | }{d \sqrt{\nu} R_{V_1}^3}  f ( t ) \;\;\;\; \text{or} \;\;\;\;  \partial_t^2 f (t) = - \frac{ | \mathbf{x}_{12} (t) | }{d \sqrt{\nu} R_{V_1}^3}  f ( t )  \, .
\end{equation}
As remarked above, the $c_n$ do not correspond to new degrees of freedom in the system---rather they reflect potential non-perturbative corrections which are not captured by the ladder expansion (we will demonstrate this explicitly in Section~\ref{sec:head-on}). In practice, we find that typically only one of the $f_n$ is consistent with the ladder expansion, and its coefficient must be fixed with an independent measurement  before the theory becomes predictive (much like a renormalization condition).

\begin{figure}
\centering
\includegraphics[width=0.95\textwidth]{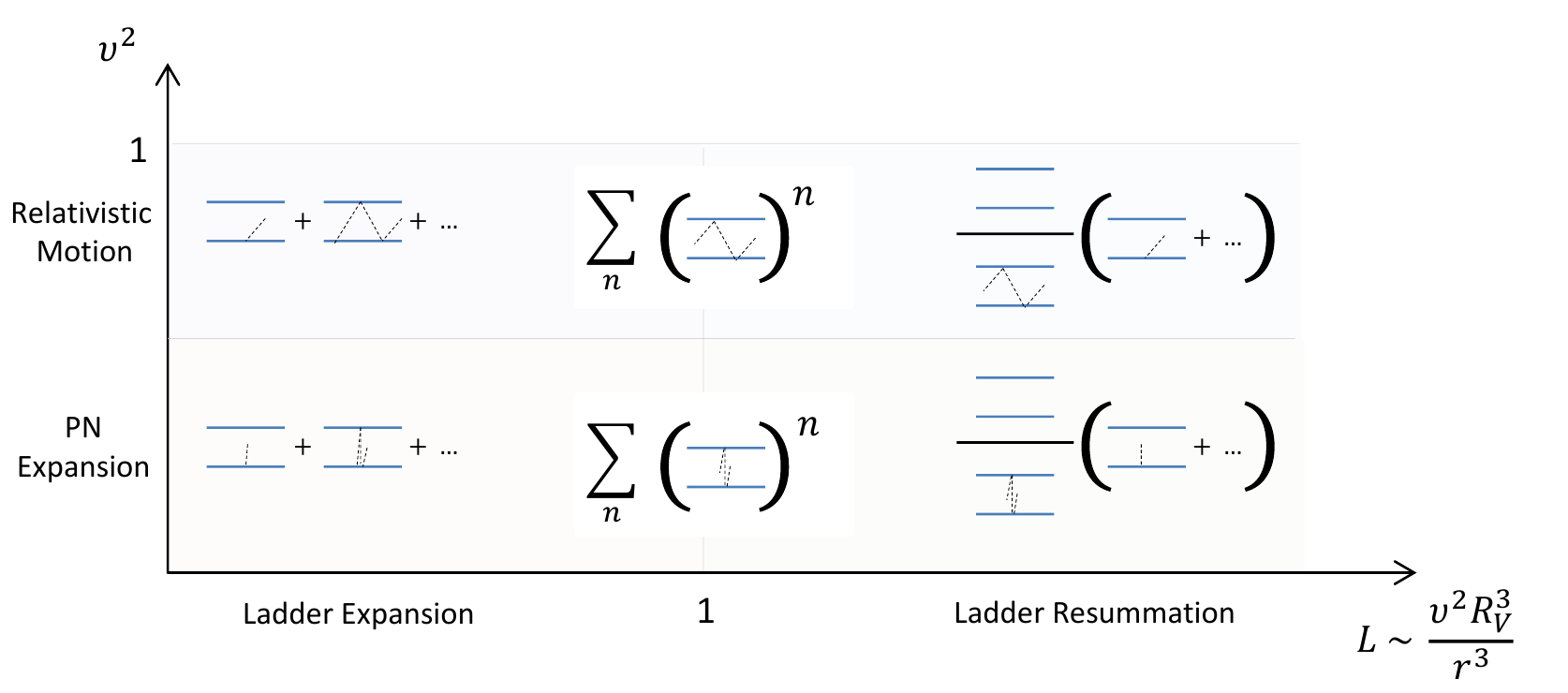}
\caption{A cartoon of the four different regimes considered in this work: the relative velocity is either relativistic ($v^2 \sim 1$) or allows for a Post-Newtonian expansion ($v^2 \ll 1$), in which case the disformal insertions can be viewed as ``instantaneous'' and the ladder diagrams lose their functional nature, becoming equal-time operators. The ladder parameter $L$ is either small enough to treat the disformal interaction perturbatively (ladder expansion), or is sufficiently large to focus on the highest derivative operators (ladder resummation). In the simplest case of a head-on collision, we will also be able to solve for intermediate values ($L \sim 1$) and show that these two ladder regimes are smoothly connected.  
\label{fig:vrladder}}
\end{figure}

\paragraph{Different Regimes:}
Deep inside the two regimes, where we can perform the ladder expansion or its inverse, solutions can be found perturbatively.
\begin{itemize}
 \item[(i)] When $r^3 \gg v^2 R_V^3$ (which is even easier to achieve in the non-relativistic limit $v \ll 1$), the ladder expansion is valid and we have found the first two terms in $G$ in \eqref{eqn:Gladder}. These (fully relativistic) expressions are valid at any velocity, but can be expanded to leading Post-Newtonian order,
\begin{align}
\frac{\varphi_1^{(0)} (t , \mathbf{x}) }{M_P}  &=    - \frac{ \con \, R_{S_1} }{ | \mathbf{x} - \mathbf{x}_1 (t) | } \left[ 1 -  \frac{ \left(  \mathbf{v}_1 (t) \cdot (  \mathbf{x} - \mathbf{x}_1 (t) ) \right)^2  }{2 | \mathbf{x} - \mathbf{x}_1 (t) |^2 } + \mathcal{O} ( \mathbf{v}_1^4 )  \right]  \, , \label{eqn:phiPN} \\
\frac{ \varphi_1^{(1)} (t , \mathbf{x}) }{M_P} &=  \frac{ \con \, \dis \, R_{S_1} R_{V_2}^3 }{ | \mathbf{x} - \mathbf{x}_1 (t) | }   \, \left[ 
\frac{  3 \left(  \mathbf{v}_{12} (t) \cdot  \mathbf{x}_{12} (t)  \right)^2 }{  | \mathbf{x}_{12} (t) |^5 }  
-  \frac{ \mathbf{a}_{12} (t) \cdot \mathbf{x}_{12} (t) }{  | \mathbf{x}_{12} (t) |^4 }  
-  \frac{ | \mathbf{v}_{12} (t) |^2    }{ | \mathbf{x}_{12} (t)  |^3 } 
+ \mathcal{O} ( \mathbf{v}_{12}^4 )    \right] ,   \nonumber
\end{align}
where $\mathbf{x}_{12} = \mathbf{x}_1 - \mathbf{x}_2$, $\mathbf{v}_{12} = \mathbf{v}_1 - \mathbf{v}_2$ and $\mathbf{a}_{12} = \mathbf{a}_1 - \mathbf{a}_2$. This agrees with the earlier results of \cite{Brax:2018bow, Brax:2019tcy}. Schematically, this means that the leading scalar fifth force in this regime is governed by $\varphi^{(0)}_1 \sim \con R_{S_1}/r \sim \con h_1^{00}$ and is comparable to the Newtonian potential, while the disformal interaction introduces small corrections, $\varphi^{(1)}_1 
\sim \dis \, v^2 R_{V_2}^3/r^3 \, \varphi^{(0)} \ll \varphi^{(0)}$.  

  \item[(ii)] For non-relativistic velocities within the Vainshtein radius, there exists a regime,
 \begin{equation}
  r^3 / R_V^3   \ll  v^2 \ll 1 
 \end{equation}
 in which the disformal terms are important, but the relativistic corrections are not. 
 \eqref{eqn:Geqn} then becomes dominated by the disformal terms in $\dis$, and a particular integral solution can be written straightforwardly,
 \begin{equation}
  \dis \, G_1^{\rm PI} (t) = - \frac{ | \mathbf{x}_{12} (t) | \, t^2 }{R_{V_1}^3} \, \sim \, \frac{r^3}{ v^2 R_V^3}   
  \label{eqn:Gsolresum}
 \end{equation}
 which is the PN expansion of \eqref{eqn:GPIeg}.
Assuming that the $c_n$ are all sufficiently small,
this gives a scalar fifth force mediated by $\varphi_1^{(0)} 
\sim \con h_1^{00} \, r^3/ \dis v^2 R_{V_1}^3 \ll h_1^{00}$, which in this regime is hugely suppressed relative to the Newtonian potential. 
We also note in passing that the possible non-perturbative corrections in this regime take the simple form,
\begin{align}
f_1 (t) \; &\propto \; \text{const}   \, , \;\;  &f_2 (t) \; &\propto \; t   \, , \;\;  \nonumber \\
f_3 (t) \; &\propto \; \int^{t} dt' \int^{t'} dt'' \, | \mathbf{x}_{12} (t'' ) |     \, , \;\; &f_4 (t) \; &\propto \; \int^{t} dt' \int^{t'} dt'' \, | \mathbf{x}_{12} (t'' ) | t''   \, .
\label{eqn:fnresumPN}
\end{align}

\end{itemize}
The ladder expansion in these different limiting cases is depicted schematically in Figure~\ref{fig:vrladder}.

\paragraph{Newtonian Ladders:}
In the Newtonian limit, the scalar signals propagate instantaneously between the two wordlines, and so physically the ladder expansion now becomes a regular series of ordinary derivative operators (rather than functional operators). In fact, the original equation of motion for $\varphi_1$ \eqref{eqn:phi1fulleom} in this limit is,
\begin{equation}
\frac{ \nabla_i \nabla^i \varphi_1 ( t, \mathbf{x} ) }{M_P} = -  \con  R_{S_1}  \delta^{(3)} ( \mathbf{x} - \mathbf{x}_1 (t) ) +  \dis \, R_{V_1}^3   \partial_t \left[  \partial_t  \varphi_2 ( t, \mathbf{x} ) \delta^{(3)} ( \mathbf{x} - \mathbf{x}_1 (t) )    \right]
\end{equation}
and so if we make the ansatz \eqref{eqn:phiansatzPN}, 
then this becomes a coupled system for the $G_A (t)$, 
\begin{align}
G_1 (t) = 1  +  \dis \, \nu R_{V_1}^3   \partial_t^2 \left(   \frac{ G_2 (t) }{ | \mathbf{x}_{12} (t) | }   \right)  \label{eqn:G1PN}  ,  \;\;\;\;
G_2 (t) = 1  +  \dis  \,  R_{V_1}^3   \partial_t^2 \left(   \frac{ G_1 (t) }{ | \mathbf{x}_{12} (t) | }   \right)  \, . 
\end{align}
Removing $G_2 (t)$ by substitution from \eqref{eqn:G1PN} then produces \eqref{eqn:Geqn}.
This explicitly demonstrates that our ladder equation is indeed equivalent to solving the original equations of motion $\varphi$.

\subsection{Effective One-Body Motion in Jordan Frame}

Deriving an effective description of the binary motion of two compact objects in General Relativity is very much an ongoing area of research. One proposal originates from Goldberger and Rothstein \cite{Goldberger:2004jt, Cardoso:2008gn, Galley:2009px} (see \cite{Foffa:2013qca, Porto:2016pyg, Levi:2018nxp} for reviews) and has recently been extended beyond GR to include the effects of a light scalar degree of freedom \cite{Endlich:2017tqa, Kuntz:2019zef, Wong:2019yoc}. Here, we are working to leading order in this EFT in which the two objects are effectively point particles, and will focus on the Newtonian force experienced due to the disformal interaction.

\paragraph{Effective One-Body Motion:}
The PN expansion of the geodesic equation \eqref{eqn:geodesic} gives the familiar Newtonian force law,
\begin{equation}
 \mathbf{a}_A^i + \Gamma^i_{\alpha\beta} u^{\alpha} u^{\beta} =  \mathbf{F}^i_{h \,A} +  \mathbf{F}^i_{\varphi \, A}
\label{eqn:geodesic_spatial}
\end{equation}
where the gravitational and fifth forces are given in \eqref{eqn:Fh} and \eqref{eqn:Fp}. For instance, the force felt by particle 2 as a result of particle 1 is, 
\begin{align}
\mathbf{F}_{h \, 2} (t) &= - \frac{ R_{S_1}  }{ | \mathbf{x}_{12} |^3 } \;\; \mathbf{x}_{12}  + \mathcal{O} \left( v^2 \frac{R_S}{r} \right)   \;\; , \\
\mathbf{F}_{\varphi \, 2}^{(\con)} (t) &= - \frac{ \con^2 }{2} G_1 \; \frac{ R_{S_1} }{ |\mathbf{x}_{12} |^3 } \; \mathbf{x}_{12}  +  \mathcal{O} \left( v^2 \frac{R_S}{r} \right)  \;\; ,  \label{eqn:FcPN}   \\
\mathbf{F}_{\varphi \, 2}^{(\dis)} (t) &=  \frac{\dis \con^2 R_{S_1} R_{V_1}^3}{ | \mathbf{x}_{12} |^4 } \, G_1  \left(   G_1 \frac{ |\mathbf{x}_{12} |^2 \partial_t \left( \mathbf{x}_{12} \cdot \mathbf{v}_{12} \right) - 3 ( \mathbf{x}_{12} \cdot \mathbf{v}_{12} )^2 }{|\mathbf{x}_{12} |^4 }   + 2 \dot{G}_1 \frac{ \mathbf{x}_{12} \cdot \mathbf{v}_{12} }{|\mathbf{x}_{12}|^2 }   - \ddot{G}_1  \right) \;\; \mathbf{x}_{12}   \label{eqn:FdPN}
\end{align}
where we have separated $F_{\varphi}$ into its conformal and disformal pieces. 

As is well known, the gravitational force obeys,
\begin{equation}
m_1 \mathbf{F}_{h \,1} ( \mathbf{x}_1 )  + m_2 \mathbf{F}_{h \, 2} ( \mathbf{x}_2 ) = 0  \;\;, \;\;\;\; 
 \mathbf{F}_{h \, 1} ( \mathbf{x}_1 ) - \mathbf{F}_{h \, 2} ( \mathbf{x}_2 ) =  \hat{\mathbf{F}}_{h} ( \mathbf{x}_{12} )
\label{eqn:EOB}
\end{equation}
where $\hat{\mathbf{F}}_{h} ( \mathbf{x}_{12} )$ is the force experienced by a particle at position $\mathbf{x}_{12}$ in the effective one-body metric\footnote{
i.e. $\hat{F}_h^\mu ( x_{12} ) = \Gamma^\mu_{\alpha\beta} u_{12}^\mu u_{12}^\nu = \Gamma^\mu_{00} + ...$ at this order, since one can think of $u_{12} = \left( 1 , \mathbf{v}_{12} \right) + ...$ as introducing a \emph{new} worldline gauge symmetry for $\mathbf{x}_{12} (t)$.  
},
\begin{equation}
\hat{h}^{00} (t , \mathbf{x} ) =  - \frac{ 2 R_S  }{ | \mathbf{x} |}   \; , \;\;\;\; \hat{h}^{ij} ( t, \mathbf{x} ) =  - \frac{ 2 R_S  }{ | \mathbf{x}  |}  \delta_{ij}    \, ,  \;\;\;\;  \hat{h}^{0i}_A ( t , \mathbf{x}  ) = 0 \, . 
\label{eqn:hEOB}
\end{equation}
where $R_S = R_{S_1} + R_{S_2}$ is the Schwarzschild scale of the total mass, $m_1 + m_2$. 
Physically, this corresponds to conservation of the total 3-momentum\footnote{
There are analogous relations for the time-like equation,
\begin{equation}
m_1 F^0_{h \,1} ( \mathbf{x}_1 )  + m_2 F^0_{h \, 2} ( \mathbf{x}_2 ) = 0  \;\;, \;\;\;\;
F^0_{h \, 1} ( \mathbf{x}_1 ) - F^0_{h \, 2} ( \mathbf{x}_2 ) =  \hat{ F}^0_{h} ( \mathbf{x}_{12} )
\end{equation}
corresponding to total energy conservation, however the geodesic equations for $x_{1}^0$ and $x_2^0$ are not independent of the gauge-fixing conditions on $\tilde{e}_1$ and $\tilde{e}_2$ and \eqref{eqn:geodesic_spatial}, so can be discarded.
}, $m_1 \mathbf{v}_1 + m_2 \mathbf{v}_2$, at leading order in $R_S/r$. The two geodesic equations are therefore most easily solved by changing variables to $m_1 \mathbf{x}_1 + m_2 \mathbf{x}_2$ and $\mathbf{x}_{12} (t)$, which reduce to the equation for a single particle moving in the effective metric \eqref{eqn:hEOB}, which is static and spherically symmetric.  
We therefore express $\mathbf{x}_{12}$ in terms of two angular variables, $\Omega$, and an effective radius, $r = | \mathbf{x}_{12} | - R_S$, which corresponds to writing the effective one-body metric in linearised Schwarzschild coordinates,
\begin{equation}
\left( \eta_{\mu\nu} +  \hat{h}_{\mu\nu} (x) \right) dx^\mu dx^\nu \Big|_{\mathbf{x} = \mathbf{x}_{12} (t)}  =  - \left( 1 - \frac{2 R_S}{r} \right) dt^2 + \left( 1 + \frac{2 R_S}{r} \right) dr^2 + r^2 d \Omega^2  \, . 
\end{equation}
at $\mathcal{O} ( R_S/ r)$. 
The resulting gravitational force from $\hat{h}_{\mu\nu}$ is then the usual Newtonian central force,
\begin{equation}
\hat{\mathbf{F}}_{h}^r = - \frac{ R_S }{r^2 } \;\; , \;\;\;\; \hat{\mathbf{F}}_{h}^\Omega =  0 \, . 
\end{equation} 

\paragraph{EOB Motion in Ladder Expansion:}
When $r^3 \gg v^2 R_V^3$, in the ladder expansion regime, we have shown that $G_A (t) = 1 + ...$ and the scalar field mediates a conformal force which is also $\mathcal{O} ( R_S/r)$, and so this modifies the effective one-body metric in the Jordan frame. 
This leading fifth force from the scalar comes entirely from the conformal piece of \eqref{eqn:Fp}, and obeys a similar relation to \eqref{eqn:EOB}, 
\begin{align}
m_1 \mathbf{F}_{\varphi \,1} ( \mathbf{x}_1 )  + m_2 \mathbf{F}_{h \, 2} ( \mathbf{x}_2 ) &= - \frac{ R_{S_1} R_{S_2} \, \mathbf{x}_{12} (t) }{ | \mathbf{x}_{12} (t)  |^3 } \left( G_1 (t) - G_2 (t)  \right)  \;\;,  \label{eqn:EOBc}  \\
 \mathbf{F}_{\varphi \, 1} ( \mathbf{x}_1 ) - \mathbf{F}_{\varphi \, 2} ( \mathbf{x}_2 ) &=   \frac{  \mathbf{x}_{12} (t) }{ | \mathbf{x}_{12} (t)  |^3 } \left( R_{S_1} G_1 (t) + R_{S_2} G_2 (t)  \right)  \nonumber
\end{align}
where the difference $G_1 - G_2 \sim \dis \, v^2 R_V^3/r^3$ is a small disformal correction. 
This means that, in the regime $r^3 \gg v^2 R_V^3$, one can construct an effective one-body Jordan-frame metric \cite{Julie:2017pkb, Julie:2017ucp},
\begin{align}
 \hat{\tilde{g}}_{\mu\nu} &= \eta_{\mu\nu} + \frac{\hat{h}_{\mu\nu} }{M_P} + \frac{ \con \; \hat{\varphi} \; \eta_{\mu\nu}  }{M_P} + ...    \\
 \text{with} \;\;\;\; \frac{ \hat{\varphi} (t, \mathbf{x} ) }{ M_P }  &= - \frac{ \con \, \left( R_{S_1} G_1 (t) + R_{S_2} G_2 (t) \right)  }{ | \mathbf{x} | }  \nonumber
\end{align}
such that the corresponding force is,
\begin{align}
\hat{\mathbf{F}} ( \mathbf{x}_{12} ) = 
\mathbf{F}_{h \, 1} ( \mathbf{x}_1 ) + \mathbf{F}_{\varphi \, 1} ( \mathbf{x}_1 ) 
-  \mathbf{F}_{h \, 1} ( \mathbf{x}_1 )  - \mathbf{F}_{\varphi \, 2} ( \mathbf{x}_2 ) \, ,
\end{align}
up to small corrections from the disformal interaction (which were computed explicitly in \cite{Brax:2018bow} in the test mass limit). This allows the geodesic equations to be reduced to an effective one-body equation for $\mathbf{x}_{12} (t)$ at leading order, just as in the usual GR problem.
If we express $\mathbf{x}_{12}$ again in terms of two angular variables, $\Omega$, and an effective radius, $r = | \mathbf{x}_{12} | - (1- \frac{\con^2}{2} ) R_S$, then we have,
\begin{equation}
\left( \eta_{\mu\nu} +  \hat{ \tilde{h} }_{\mu\nu} (x) \right) dx^\mu dx^\nu \Big|_{\mathbf{x} = \mathbf{x}_{12} (t)}  =  - \left( 1 - \frac{ (2 + \con^2) R_S}{r}  \right) dt^2 + \left( 1 + \frac{ (2 - \con^2) R_S}{r}  \right) dr^2 + r^2 d \Omega^2  \, . 
\end{equation}
up to $\mathcal{O} ( R_S^2 / r^2 )$. 
The effective force is then the usual Newtonian central force plus a modification from the scalar conformal coupling,
\begin{equation}
\hat{\mathbf{F}}^r = - \frac{ R_S }{r^2} \left( 1 + \frac{\con^2}{2}  \right)  \;\; , \;\;\;\; \hat{\mathbf{F}}^\Omega =  0 \, . 
\end{equation} 
which has the effect of rescaling Newton's constant by a factor of $\left( 1 + \con^2/2 \right)$. 

The vanishing of the angular force corresponds to the angular momentum, $\mathbf{J}$, being conserved, and 
the angular geodesic equations are then easily solved,
\begin{equation}
 r^2 (t) \dot{\theta} (t)  = J  \;\; , \;\;\;\;  
 \label{eqn:J}
\end{equation}
where we have written $\Omega$ in terms of two spherical angles, one of which (the angle between $\mathbf{x}_{12}$ and $\mathbf{J}$) is not dynamical, and the other ($\theta$) rotates in the plane orthogonal to $\mathbf{J}$. 
The radial geodesic equation is then, 
\begin{equation}
 \ddot{r} - \frac{J^2}{r^3} = - \frac{R_S}{r^2} \left(  1 + \frac{\con^2}{2}  \right) + \mathcal{O} \left(  \frac{R_S^2}{r^2} , \; \dis \frac{v^2 R_V^3}{r^3}    \right)  \; , 
 \label{eqn:reomexp}
\end{equation}
at leading order. The subleading disformal corrections enter in two ways: through an explicit force term $\mathbf{F}_{\varphi}^{(\dis)}$ (shown in \eqref{eqn:Fp}), but also through an induced coupling between $\mathbf{x}_{12} (t)$ and the center of mass motion (according to \eqref{eqn:EOBc}). This leads to small perturbations to the Newtonian motion, which can be even larger than the relativistic corrections of GR if $\dis v^2 R_V^3 /r^3  \gg R_S^2/r^2$ (equivalent to $ \dis  r / R_S \gg  \, H_0^2 t^2$ for typical dark energy scales). 

\paragraph{EOB Motion with Ladder Resummation:}
However, in the ladder resummation regime, the disformal piece of the fifth force can dominate. Unless $G$ is sufficiently screened, the disformal fifth force is much larger than both the gravitational and the conformal force. This is problematic, since the disformal force does \emph{not} obey a relation like \eqref{eqn:EOB}---this can be understood as the disformal interaction modifying the particle momenta,
\begin{align}
 p_{A \, \mu} &= m_A u_{A\, \mu}  + \frac{\dis}{\M^4} \, \frac{\partial \varphi (x_A)}{\partial x_A^\mu } \, \partial_\tau \varphi ( x_A ) + ...  \\
 \Rightarrow \;\;\;\; \mathbf{p}_A &\sim  m_A \mathbf{v}_A \left( 1 +  \dis \nu^2 G^2 \frac{R_S R_V^3}{ | \mathbf{x}_{12}|^4}   \right)   \nonumber
\end{align}
and so unless\footnote{
or $\dis \to 0$ or $\nu \to 0$, but these limits correspond to turning off the disformal interaction altogether and the motion reduces to that of a conformally coupled scalar.
} $ G^2 \ll r^4/R_S R_V^3$ when $r^4/R_S R_V^3 \ll 1$, then the particle momenta receives large corrections which invalidates the decoupling between center of mass motion $(m_1 \mathbf{x}_1 + m_2 \mathbf{x}_2)$ and relative motion $(\mathbf{x}_1 - \mathbf{x}_2)$, even at very low (naively Newtonian) velocities. Even if one could devise a new effective one-body scheme for this case, it would clearly be incompatible with everyday observations.
We will assume that {\bf the ladder resummation screens the scalar field profile}, so that these disformal forces are subdominant to the usual gravitational force (and $\tilde{g}_{\mu\nu} = g_{\mu\nu} +$ small corrections throughout the motion), which will be justified a posteriori by the solutions that we find to \eqref{eqn:Geqn}. 
This rescues a simple effective one-body description in the Newtonian limit, 
\begin{equation}
 \ddot{r} - \frac{J^2}{r^3} = - \frac{R_S}{r^2}  + \mathcal{O} \left(  \frac{R_S^2}{r^2} , \; \con \frac{R_S}{r} G    ,  \; \dis \frac{ R_S R_V^3 }{ r^4 } G^2  \right)
 \label{eqn:reomresum}
\end{equation}
in which both the conformal and disformal fifth forces are comparable, and provide small corrections to the Newtonian trajectory. 
Once we have solved this effective one-body problem for $\mathbf{x}_{12} (t)$, we can then use this to compute $G(t)$ using \eqref{eqn:Geqn} (c.f. \eqref{eqn:Gsolresum}) and show that it indeed exhibits screening---hence justifying our initial assumption.

\paragraph{Hard vs. Soft Scattering:}
In both regimes, $r^3 \gg v^2 R_V^3$ and $r^3 \ll v^2 R_V^3$, the leading order geodesic equation (\eqref{eqn:reomexp} or \eqref{eqn:reomresum}) corresponds to that of linearised Schwarzschild. It can be integrated once, 
\begin{equation}
 E = \frac{1}{2} \dot r^2 +  \frac{ J^2 }{2 r^2} -  \frac{R_S'}{r}  \, . 
 \label{eqn:Eeom}
\end{equation}
where $E$ and $J$ are constants of motion (the energy and angular momentum respectively), and $R_S'$ represents $R_S ( 1 + \con^2/2)$ in the ladder expansion regime ($L \ll 1$) and $R_S$ in the ladder resummation regime ($ L \gg 1 $). 
For \emph{hard} scattering (large $\dot{r}$ or large $r \dot{\theta}$), the term in $R_S'$ is a small correction and the motion is approximately inertial. On the other hand, for small velocities,
\begin{equation}
 \dot{r} \sim \sqrt{ \frac{R_S'}{r} }  \, , \;\;\;\; J \sim \sqrt{R_S' r}    \, . 
\end{equation}
the motion becomes bound ($E < 0$), and the gravitational force provides an order one effect to the trajectory. For hard scattering, the two bodies are initially far apart ($L \ll 1$) and a ladder expansion is valid, but as they approach each other it breaks down ($L \gg 1$, for sufficiently small impact parameter) and the scalar field smoothly transitions to a screened solution. For bound orbits, on the other hand, the bodies can remain screened (in the $L \gg 1$ regime) throughout their periodic motion, in which case the disformally coupled scalar is only ever a small perturbation.  
We will demonstrate this behaviour explicitly in the following section.

\section{Examples of Ladder Screening}
\label{sec:eg}

In this section, we present various examples of our ladder resummation and screening mechanism in action. 
First, we will focus on hard scattering with relative particle velocity $R_S/r \ll v^2 \ll 1$. In this case, the particles move inertially towards each other from an initially large separation, pass through an intermediate region (where the ladder resummation kicks in and provides screening), and then go out again to large separations (where Newtonian $\varphi \sim 1/r$ behaviour is recovered). This is shown in Figure~\ref{fig:flyby}, and demonstrates that it is possible to smoothly transition from a Newtonian ladder expansion regime to the screened regime. 
Then, we will turn to initial particle velocities $v^2 \sim R_S/r$, for which gravitationally bound orbits form. If the orbital separation is well within the ladder resummation regime, then we can compute $G_1 (t)$ and show that it is very efficiently screened\footnote{
One caveat to this is the non-perturbative corrections captured by the $c_n$ in \eqref{eqn:Gsol}: we tacitly assume that $c_n$ can be fixed by the condition that $\varphi$ and its derivatives are initially very small, and then show that $\varphi$ will remain small for many orbits. 
Since the orbit never leaves the ladder resummation regime, we are not able to match $c_n$ onto the asymptotic boundary conditions, as we are able to for the hard scattering case.
}, as depicted in Figure~\ref{fig:screening}. 
These results, together with solar system observations of planetary orbits, will allow us to place constraints on the $(\con, \dis)$ parameter space in Section~\ref{sec:solarsystem}, which for the first time properly account for disformal effects at small separations.

\subsection{Head-On Collision}
\label{sec:head-on}

First we will consider hard scattering ($v^2 \gg R_S/r$), so the trajectories are approximately inertial (at leading order in $R_S/r$). 
The most symmetric case is when the two particles collide head-on, so that the motion takes place along a line in space. For instance, in the rest frame of particle 1, the trajectories are, 
\begin{equation}
  \mathbf{x}_1 ( t ) = \left( \begin{array}{c} 
 0  \\
 0 \\
 0
 \end{array} \right)  \; , \;\;\;\;  \mathbf{x}_2 ( t ) = \left( \begin{array}{c} 
 0  \\
 0 \\
 v t
 \end{array} \right)
 \label{eqn:head-on}
\end{equation}
for constant relative velocity $v$.

In terms of the dimensionless variable,
\begin{equation}
\z =  \frac{ v t }{  v^{2/3} R_{V_1} } ( \dis^2 \nu )^{1/6}  \, ,  
\end{equation}
\eqref{eqn:Geqn} takes on a particularly simple form:  
\begin{align}
G_1 - \frac{\partial^2}{\partial \z^2} \left( \frac{1}{\z}  \frac{\partial^2}{\partial \z^2} \left( \frac{ G_1 }{ \z } \right) \right)  = 1 +  \sqrt{\nu} \frac{\partial^2}{\partial \z^2} \left( \frac{1}{\z}  \right) \, .
\label{eqn:Geqnlineb0}
\end{align}
where we have restricted our attention to $\z > 0$ (as we can later use the reflection $G_1 ( \z ) = G_1 ( - \z)$ to construct the whole solution). Note that $\sqrt{\nu}$ may have either sign (corresponding to whether $\dis >0$ or $\dis <0$).

\paragraph{Ladder Expansion:}
When $\z \gg 1$, we can perform a ladder expansion, 
\begin{equation}
G_1  (t) = 1 + \sqrt{\nu} \frac{ 2 }{ \z^3 } + \mathcal{O} \left( \frac{1}{\z^6}  \right)  \, .  \;\;\;\;
\label{eqn:GheadonL}
\end{equation}
Since the derivative operators in \eqref{eqn:Geqnlineb0} are suppressed, there are no integration constants to contend with in this regime---the ladder expansion is unique. 

The corresponding force experienced by particle 2 from this $\varphi_1 (t, \mathbf{x} ) = - \con R_{S_1} G_1 (t) / | \mathbf{x} |$ profile sourced by the two-body motion is,
\begin{align}
 \mathbf{F}_{\varphi \, 2}^{(\con)} (t) &= \frac{\con^2}{2} \left( 1 + \sqrt{\nu} \frac{ 2 }{ \z^3  }  + ... \right)    \mathbf{F}_{h \, 2} (t)  \; ,  \nonumber \\
  \mathbf{F}_{\varphi \, 2}^{(\dis)} (t) &=  \dis \con^2 \left( \sqrt{\nu} \frac{ 2 }{  \z^3 } + ... \right) \mathbf{F}_{h \, 2} (t) \; ,  \label{eqn:FLadderHeadOn}
\end{align}
where $\mathbf{F}_{h \, 2} (t)$ is the usual gravitational force.
The disformal interaction provides a small correction $( \dis/\z^3 \sim  v^2 R_V^3 / (vt)^3 \ll 1 )$, but the leading conformal force is the same order as $\mathbf{F}_h$ (when $\con$ is order unity). 

\paragraph{Solving the Ladder Equation:}
Let us first consider the homogeneous part of \eqref{eqn:Geqnlineb0}, namely,
\begin{equation}
\frac{G_1}{\z} = \frac{1}{\z} \frac{\partial^2}{\partial \z^2} \left( \frac{1}{\z}  \frac{\partial^2}{\partial \z^2} \left( \frac{ G_1 }{ \z } \right) \right) \, , 
\end{equation}
whose solutions are nothing more than the four Airy functions, which have asymptotic expansions,
\begin{align}
\Ai (z) &\sim  \frac{ e^{- 2 z^{3/2}/3 } }{z^{1/4}}    \,\, , \;\;\;\; &\Bi (z) &\sim  \frac{ e^{2 z^{3/2}/3 } }{z^{1/4}}    \nonumber \\
\Ai ( -z) &\sim \frac{ \sin ( \frac{\pi}{4} + \frac{ 2 z^{3/2}}{3} ) }{z^{1/4}}   \,\, , \;\;\;\; &\Bi ( -z) &\sim  \frac{ \cos ( \frac{\pi}{4} +  \frac{2 z^{3/2}}{3} ) }{z^{1/4}}   \, .   \label{eqn:Airy}
\end{align}
As we can see, it is only $\Ai ( \z )$ (which decays at large $\z$) which can be freely added to $G_1$, the coefficients of the other three complementary functions must be carefully fixed in order to respect the ladder expansion \eqref{eqn:GheadonL}.   

The most general solution to \eqref{eqn:Geqnlineb0} consistent with the ladder expansion is found to be,
\begin{align}
\frac{ G_1 }{  \zz } = \frac{\pi}{2}  \Bigg[ c_1 \Ai (\zz ) &+ 
( 1 + \sqrt{\nu} ) \left(   
\frac{1}{3} \Bi  ( \zz ) +  \Ai (\zz) \IBi (\zz)  - \Bi (\zz) \IAi ( \zz )
 \right)
 \nonumber \\
& + ( 1 - \sqrt{\nu} ) \left(
\frac{2}{3} \Bi  ( - \zz ) -  \Ai (- \zz ) \IBi (- \zz ) +  \Bi (- \zz )  \IAi (- \zz )
 \right) \Bigg]  
 \label{eqn:Gsolb0}
\end{align}
which contains one undetermined constant ($c_1$), and where $\IAi$ and $\IBi$ are the integrals $ \IAi (z ) = \int d z  \, \Ai ( z ) \, $ and $ \IBi (z ) = \int d z  \, \Bi ( z ) \, $. 
This solution is valid for all $\z$ (at leading order in $v^2$ and $R_S/r$), and in particular if we expand at large separations, 
\begin{equation}
 \lim_{\z \gg 1} G_1  = 1  + \sqrt{\nu} \frac{2}{ \z^3} + ... 
\end{equation}
we recover the ladder expansion. \eqref{eqn:Gsolb0} therefore represents the complete resummation of the ladder series. The undetermined coefficient arises as a \emph{non-perturbative} correction, since at large $\z$, $ \Ai ( \z ) \sim   \exp \left( -  \frac{2 \z^{3/2} }{3}  \right) /  \z^{1/4}  $ and cannot be seen in a Taylor series expansion about $\z = \infty$ at any order in perturbation theory.

\paragraph{Ladder Screening:}
At small $\z$, the ladder diagrams have resummed into,
 \begin{equation}
\lim_{ \zz \ll 1 } G_1 = 
 - a_1 \zz  
 - a_2 \zz^2  
 - a_3 \zz^3     
 - a_4 \zz^4  
  - a_5 \zz^5  
  - a_6 \zz^6  +  \mathcal{O} ( \zz^7 ) \, . 
  \label{eqn:Gsolb0lowz}
\end{equation}
where,
\begin{align}
 a_1 &= \pi \frac{ \bar{c}_1 }{3^{+1/6} \Gamma ( - \tfrac{1}{3}) } \;\;, \;\; &a_2 &= \frac{\pi}{3} \frac{  \bar{c}_1 - 1 - \sqrt{\nu} }{3^{-1/6} \Gamma (\tfrac{1}{3} )}  \;\;, \;\;  &a_3 &= \frac{ \sqrt{\nu} }{2} \;\;, \;\;  \nonumber \\
 a_4 &= \frac{ \pi}{ 6}   \frac{ \bar{c}_1 - 2 + 2 \sqrt{\nu}   }{ 3^{1/6} \Gamma ( - \tfrac{1}{3} ) } \;\; , \;\; &a_5 &= \frac{ \pi}{ 36 }   \frac{ \bar{c}_1 - 3 + \sqrt{\nu}   }{ 3^{-1/6} \Gamma ( \tfrac{1}{3} ) }  \;\;, \;\; &a_6 &= \frac{1}{40} \,  .   \label{eqn:adef}
\end{align}
and $\bar{c}_1 = \tfrac{\sqrt{3}}{\pi} c_1 + \tfrac{1}{2} \left( 3 - \sqrt{\nu} \right) \pi$. 
The coefficients of all $\z^{3n}$ terms in this expansion are uniquely fixed by the perturbative ladder resummation, while the $\z^{3n+1}$ and $\z^{3n+2}$ terms are sensitive to non-perturbative effects. 
Note that it is \emph{not} possible to fix $\bar{c}_1$ by, e.g., demanding a smooth field profile at $\z = 0$ (but this is perhaps not surprising since in this toy one-dimensional example the Newtonian force $\mathbf{F}_{h \, 2}$ is also not smooth when the particles collide)---rather $\bar{c}_1$ should be viewed as a renormalisation constant which must be fixed (using a single measurement) before the theory becomes predictive.   

This scalar field profile generated as two particle fly past each other demonstrates that screening can take place as a result of the resummation of ladder diagrams. 
If one considers the force experienced\footnote{
As discussed above, this result does not immediately apply to light deflection experiments such as those carried out by \emph{Cassini} (and which provide strong bounds on the PPN parameters $\gamma$), because the massless photon can not be treated non-relativistically. Further analysis of the fully relativistic \eqref{eqn:GeqnRela} is required to assess whether a disformally coupled scalar is compatible with solar system tests involving the motion of photons. 
} by particle 2 in the regime $\z \ll \b$,
\begin{align}
 \mathbf{F}_{\varphi \, 2}^{(\con)} (t) &=  - \frac{\con^2}{4 \dis} \left( \z^3 + ... \right)    \mathbf{F}_{h \, 2} (t)  \; , \\
  \mathbf{F}_{\varphi \, 2}^{(\dis)} (t) &=  \frac{2 \con^2}{ \dis} \z \left(
  \frac{ \dis^2 a_1 a_3 }{ (\dis^2 \nu )^{2/3} } 
  + \frac{ \dis^2 \z ( a_2 a_3 + 3 a_1 a_4 ) }{ ( \dis^2 \nu )^{5/6} } 
  + \frac{ \z^2 ( a_3^2 + 3 a_2 a_4 + 6 a_1 a_5 ) }{ \nu } + ... \right) \mathbf{F}_{h \, 2} (t) \; ,
\end{align}
we find that it is indeed significantly smaller than the Newtonian gravitational force,
suppressed by at least one factor of $\hat{z} \sim  r^3 / v^3 R_{V_1}^3 $, where $r$ is the separation of the particles.  

This result provides an important proof of concept---that it is possible to have screened solutions in the interior which smoothly connect to the ladder expansion at large separations. It also demonstrates how non-perturbative corrections can arise at short distances whose coefficients must be calibrated before the theory becomes fully predictive. 
We will now discuss these constants of integration in more detail, form the point of view of resurgence.

\paragraph{Borel resummation:}
There is another (equivalent) way to resum the ladder expansion, which sheds some light on the undetermined integration constant, $c_1$. 
In this simple case, the ladder expansion can be expressed as the infinite series,
\begin{align}
 G_1 
 &= \sum_{n=0}^{\infty}  \frac{ (6n)!}{ (6n)!!!} \frac{1}{\z^{6n}} + \sqrt{\nu} \sum_{n=0}^{\infty}  \frac{ (6n+3)!}{ (6n+3)!!!} \frac{1}{\z^{6n+3}}
\end{align}
where we have used a multifactorial\footnote{
This can also be written in terms of the Gamma function as $(3z)!!!= 3^{z} \Gamma \left(1 + z \right) $.
}, $n!!! = n (n-3) (n-6) ...$, with the convention that $0!!! = 1$. 

This is an \emph{asymptotic} series, including more terms only improves the accuracy of the expansion if $\z$ is sufficiently large\footnote{
At any finite $\z$, there is an optimal (finite) number of terms to include in this series (which gives a \emph{superasymptotic} approximation with bounded errors).
} (and the series as a whole only converges if $\z$ is formally infinite).  
But note that the coefficients of $\z^{-n}$ in this series are bounded by
$n!$, and so the following \emph{Borel transform} of the sum,
\begin{equation}
 \mathcal{B} \left[ \sum_n a_n z^{-n-1}   ,  \, w  \right] = \sum_n \frac{a_n}{n!} w^n  
\end{equation}
\emph{does} converge. Using the identity,
\begin{equation}
 1 = \frac{1}{n!} \int_0^\infty d w \, w^n e^{-w} 
\end{equation}
we see that the Borel transform can be undone by a Laplace transform\footnote{
The sum and the integral can be interchanged since the Borel transformed sum now converges.
}, and so one resummed expression for $G_1$ is,
\begin{equation}
 \frac{G_1^B (\z)}{\z} = \int_0^{\infty} d w \, e^{- \z w } \left(  \mathcal{B} \left[ \sum_{n=0}^{\infty}  \frac{ (6n)!}{ (6n)!!!} \frac{1}{\z^{6n+1}} , w \right] + \sqrt{\nu} \mathcal{B} \left[ \sum_{n=0}^{\infty}  \frac{ (6n+3)!}{ (6n+3)!!!} \frac{1}{\z^{6n+4}} , w \right]  \right) \, .  
 \label{eqn:GBorel}
\end{equation}
This expression is known as the \emph{Borel resummation} of the series\footnote{
and gives a \emph{hyperasymptotic} approximation with further improved errors
}. See \cite{Boyd1999} for a review of these concepts.

\paragraph{Stokes phenomena:}
The Borel transform of a series can contain singularities at particular values of $w$. This leads to an ambiguity in how the Laplace transform is taken, and corresponds to possible non-perturbative corrections to the series.  
This is best illustrated with a simple example. Consider the asymptotic series, 
\begin{equation}
 S (z) = \sum_n n! z^{-n-1}  \, . 
 \label{eqn:Euler_eg}
\end{equation}
The Borel transform is given by,
\begin{equation}
 \mathcal{B} [ S (z) , \, w ] = \sum_n w^n  = \frac{1}{1-w}
\end{equation}
which has a pole at $w = 1$. This complicates the Laplace transform needed to undo $\mathcal{B}$, since we need to specify whether the integration contour should go over or under the pole. The difference between the two contours is given by the residue of the pole in $\mathcal{B}$,
\begin{equation}
 S^B_+ (z) - S^B_- (z) =  \oint_1 d w \, \frac{ e^{-z w} }{1 - w} = 2 \pi i e^{-w} \, . 
\end{equation}
This is known as a \emph{Stokes discontinuity} (and the line $\int_0^{\infty} d w$ is known as a \emph{Stokes line}), and corresponds to an \emph{ambiguity} in the Borel resummation. There are two different resummations we can associate to \eqref{eqn:Euler_eg}, namely
\begin{equation}
 S^B_{\pm} (z) = e^{-z} \text{Ei} (z)  \pm i \pi e^{-z}
 \label{eqn:Borel_eg}
\end{equation}
When expanded at large $z$, both of these return the original series plus a non-perturbative piece\footnote{
Retaining non-perturbative pieces in this way is known as a \emph{trans-series}. 
}, 
\begin{equation}
\lim_{z \gg 1} S^B_{\pm} (z) =  \sum_n  n! z^{-n-1} \pm i \pi e^{-z} \,  . 
\end{equation}
The non-perturbative part is naively invisible to the perturbative expansion, but thanks to the Stokes discontinuity we can actually recover it from the perturbative resummation. This is the underlying principle of \emph{resurgence}---see \cite{Dorigoni:2014hea} for an introduction to the subject. 

In our approach to the ladder resummation, we instead resum the series by inverting a differential operator. In the simple example \eqref{eqn:Euler_eg}, this corresponds to,
\begin{align}
 S (z) = \sum_n  (-1)^n \partial_z^n \left( \frac{1}{z} \right)  &= \frac{1}{1 + \partial_z} \left( \frac{1}{z} \right)  \nonumber \\
\Rightarrow \;\; (1 + \partial_z ) S (z) &= \frac{1}{z}  \, . 
\end{align}
This first order differential equation has a general solution with one integration constant,
\begin{equation}
 S (z) =  e^{-z} \text{Ei} (z)   + c_1 \, e^{-z}
\end{equation}
and we immediately recognise this integration constant as the ambiguity in the Borel resummation \eqref{eqn:Borel_eg}. 

Note that while for simple trajectories like \eqref{eqn:head-on} we can perform the Borel resummation exactly, in general we have only the differential equation \eqref{eqn:GeqnRela}. But the interpretation of these integration constants is now clear---they correspond to ambiguities encountered when resumming this asymptotic series.

\paragraph{A Disformal Instanton:}
Returning to our ladder series, the Borel transforms are explicitly,
\begin{align}
 \mathcal{B} \left[ \sum_{n=0}^{\infty}  \frac{ (6n)!}{ (6n)!!!} \frac{1}{ \z^{6n+1}} , w \right]  &=  \text{cosh} \left( \frac{w^3}{3} \right)  \, ,   \\
 \mathcal{B} \left[ \sum_{n=0}^{\infty}  \frac{ (6n+3)!}{ (6n+3)!!!} \frac{1}{ \z^{6n+4}} , w \right] &= \text{sinh} \left( \frac{w^3}{3} \right)  \, . 
\end{align}
Both of these have singularities at $w = \infty$, and so the forward Laplace transform \eqref{eqn:GBorel} does not converge unless the integration contour is deformed. The different choices of contour correspond to an ambiguity in $G^B_1 ( \z)/\z$, and is nothing more than the $c_1 \, \Ai ( \z )$ complementary function we encountered above.  

One way to view this non-perturbative correction is as coming from an \emph{instanton} of the effective action which governs the $\varphi_1 \left[ x_1 (\tau ) \right]$ functional. That is, using the Borel resummation we can express the field profile using a ``path integral'',
\begin{equation}
 \frac{ G_1^B ( \z ) }{ \z } =  \frac{ 1+ \sqrt{\nu} }{2} \int_0^{\infty} d w \, e^{S_{\rm eff}^{+} (w, \z) } + \frac{ 1 - \sqrt{\nu} }{2} \int_0^{\infty} d w \, e^{S_{\rm eff}^- (w, \z )}
\end{equation}
where the effective actions are,
\begin{equation}
 S_{\rm eff}^{\pm} (w, \z ) = - \z w \pm \frac{w^3}{3} \, . 
\end{equation}
The saddle points of $S_{\rm eff}^{+}$ are,
\begin{equation}
w= \pm \sqrt{\z}  \;\;\;\; \text{at which} \;\;\;\; S_{\rm eff}^{+} = \mp \frac{2}{3} \z^{3/2}
\end{equation}
and the saddle points of $S_{\rm eff}^{-}$ are,
\begin{equation}
w = \pm i \sqrt{\z} \;\;\;\; \text{at which} \;\;\;\; S_{\rm eff}^{-} = \mp \frac{2 i}{3} \z^{3/2}  \, . 
\end{equation}
Each of these saddle points corresponds to the asymptoptic behaviour of the four Airy complementary functions \eqref{eqn:Airy} which appeared in our general solution to \eqref{eqn:Geqnlineb0}.

These different saddle points (different solutions to the equations of motion) in the QFT context are called instantons, and provide non-perturbative corrections like those found above. 
One of the main ideas behind resurgence is to learn about these instantons (saddle points in an effective action) using the ambiguities (discontinuities across Stokes lines) in the perturbative resummation.
Here we see that a similar philosophy can be applied to two-body systems, where the ladder resummation can be used to study the non-perturbative corrections which arise at short distances.

\subsection{Rutherford Scattering}
\label{sec:flyby}

Now consider the more general scattering problem, in which the particles pass each other with impact parameter $b \neq 0$. 
Without loss of generality, we can orient our coordinate system so that,
\begin{equation}
  \mathbf{x}_1 ( t ) = \left( \begin{array}{c} 
 0  \\
 0 \\
 0
 \end{array} \right)  \; , \;\;\;\;  \mathbf{x}_2 ( t ) = \left( \begin{array}{c} 
 0  \\
 b \\
 v t
 \end{array} \right)
\end{equation}
for constant relative velocity $v$. 

Introducing the dimensionless variables,
\begin{equation}
 \b = \frac{b}{ v^{2/3} R_{V_1} }  ( \dis^2 \nu )^{1/6}   \;\; , \;\;\;\; \z =  \frac{ v t }{  v^{2/3} R_{V_1} }  ( \dis^2 \nu )^{1/6} 
\end{equation}
then \eqref{eqn:Geqn} becomes:  
\begin{align}
G_1 -    \frac{\partial^2}{\partial \z^2} \left( \frac{1}{\sqrt{ \b^2 + \z^2} }  \frac{\partial^2}{\partial \z^2} \left( \frac{ G_1 }{\sqrt{ \b^2 + \z^2} } \right) \right)  = 1 +  \sqrt{\nu} \frac{\partial^2}{\partial \z^2} \left( \frac{1}{\sqrt{ \b^2 + \z^2} } \right) \, .
\label{eqn:Geqnline}
\end{align}
Since $\z$ ranges from $-\infty$ to $+\infty$, the particles cannot be in the ladder resummation regime ($ \z^3 \ll 1 $) for the entire trajectory. We distinguish instead between large impact parameters ($b^3 \gg v^2 R_V^3$), for which the ladder expansion is valid for the whole wordline, and small impact parameters ($b^3 \ll v^2 R_V^3$), for which the ladder expansion breaks down at some intermediate time. 

\paragraph{Ladder Expansion:}
When $\b$ is very large, we can construct perturbative solutions for $G_1 (t)$ in two regions,
\begin{align}
 G_1^{\rm out} (t) &= 1 + \sqrt{\nu}  \frac{2}{\z^3} + \mathcal{O} \left( \frac{1}{\z^6}  \right)  \;\;\;\; &&\text{when } \z  \gg \b   \\
 G_1^{\rm in} (t) &= 1 - \sqrt{\nu} \frac{ \b^2 - 2 \hat{z}^2 }{ ( \b^2 + \z^2 )^{5/2} } + \mathcal{O} \left( \frac{1}{\b^6}  \right)   \;\;\;\; &&\text{when } \z \lesssim \b
\end{align}
Since the derivative operators in \eqref{eqn:Geqnline} are always suppressed by large $\b$, there are no integration constants to contend with in either regime. The matching between the two asymptotic expansions is then trivially satisfied, namely $\lim_{\z \ll \b} G_1^{\rm out} (t) = \lim_{\z \gg \b} G_1^{\rm in} (t)$. 

At large separations, when $\z \gg \b$, the force experienced by particle 2 from this profile is the same as the head-on collision case \eqref{eqn:FLadderHeadOn}, and the impact parameter is unimportant. Near the distance of closest approach, when $\z \lesssim \b$, the impact parameter modifies the small disformal corrections to, 
\begin{align}
 \mathbf{F}_{\varphi \, 2}^{(\con)} (t) &= \frac{\con^2}{2} \left( 1 - \dis \frac{ \b^2 - 2 \hat{z}^2 }{ ( \b^2 + \z^2 )^{5/2} }  + ... \right)    \mathbf{F}_{h \, 2} (t)  \; , \nonumber \\
  \mathbf{F}_{\varphi \, 2}^{(\dis)} (t) &= - \dis \con^2 \left( \frac{ \b^2 - 2 \z^2}{  ( \b^2 + \z^2 )^{5/2} } + ... \right) \mathbf{F}_{h \, 2} (t) \; ,
\end{align}
where $\mathbf{F}_{h \, 2} (t)$ is the usual gravitational force.
The disformal interaction provides a small correction $( \dis/\b^3 \sim \dis v^2 R_V^3 / b^3 \ll 1 )$, but the leading conformal force remains the same order as $\mathbf{F}_h$ (when $\con$ is order unity).

\paragraph{Ladder Resummation:}
Now consider small impact parameters, $\b \ll 1$. We can again solve for $G_1$ perturbatively in two regimes,
\begin{align}
G_1^{\rm out} - \frac{\partial^2}{\partial \z^2} \left( \frac{1}{\z}  \frac{\partial^2}{\partial \z^2} \left( \frac{ G_1^{\rm out} }{ \z } \right) \right)  &= 1 +  \sqrt{\nu} \frac{\partial^2}{\partial \z^2} \left( \frac{1}{\z}  \right) + \mathcal{O} \left( \b^2 \right) \, .
  \,   \;\;\;\; &&\text{when  } \z \gg \b   \nonumber  \\
 \frac{\partial^2}{\partial \z^2} \left( \frac{1}{\sqrt{ \b^2 + \z^2 }}  \frac{\partial^2}{\partial \z^2} \left( \frac{ G_1^{\rm in} }{ \sqrt{ \b^2 + \z^2 } } \right) \right)  &= - \sqrt{\nu} \frac{\partial^2}{\partial \z^2} \left( \frac{1}{ \sqrt{ \b^2 + \z^2 } }  \right) + \mathcal{O} \left( \b^2 \right) \,    \;\;\;\; &&\text{when  } \z \lesssim \b    
\end{align}
Note that when $\z \gg \b$, we have approximately the same equation as the head-on collision case \eqref{eqn:Geqnlineb0}, whose solution is given by \eqref{eqn:Gsolb0}. In the inner region, where $\z \lesssim \b$, we have a new solution,
\begin{equation}
\frac{ G_1^{\rm in} }{ \sqrt{ \b^2 + \z^2 } } =  - \frac{\sqrt{\nu}}{2} \z^2 + \sum_{n=1}^4 d_n \, g_n (z)
\label{eqn:Ginb}
\end{equation}
where now the complementary functions are,
\begin{align}
g_1 ( \hat{z} ) &= 1   , \;\;\;\; g_2 (\hat{z} ) =  \hat{z}   \nonumber  \\
g_3 ( \hat{z} ) &= 3 \b^2 \z \log \left( \z + \sqrt{\z^2+\b^2} \right)  + (\z^2 - 2 \b^2 )  \sqrt{ \z^2 + \b^2}   , \;\;\;\;  \nonumber \\
g_4 (\hat{z} ) &=   \frac{3}{2} \b^4 \log \left(  \z + \sqrt{\z^2+\b^2}  \right) + \left(  \z^2 + \frac{5}{2} \b^2  \right) \z \sqrt{ \z^2 + \b^2}  \, , 
\end{align}
at this order.
When we perform the matching, $\lim_{\z \ll \b} G_1^{\rm out} (t) = \lim_{\z \gg \b} G_1^{\rm in} (t)$, the four $d_n$ map onto the four $c_n$ of \eqref{eqn:Gsol} (with $f_n$ given by \eqref{eqn:Airy}), and then consistency with the ladder expansion removes all but one: explicitly, comparing with \eqref{eqn:Gsolb0lowz}, 
\begin{align}
 d_1 = -a_1  \;\; , \;\; d_2 = -a_2 \;\;, \;\; d_3 = -a_4 \;\; , \;\; d_4 = -a_5 \;\; ,
\end{align} 
and $c_1$ remains the only undetermined coefficient. Taken together, \eqref{eqn:Ginb} for $G^{\rm in}$ and \eqref{eqn:Gsolb0} for $G^{\rm out}$ give a complete approximation for the scalar field profile at all $\z$. 

\begin{figure}
\includegraphics[width=0.42\textwidth]{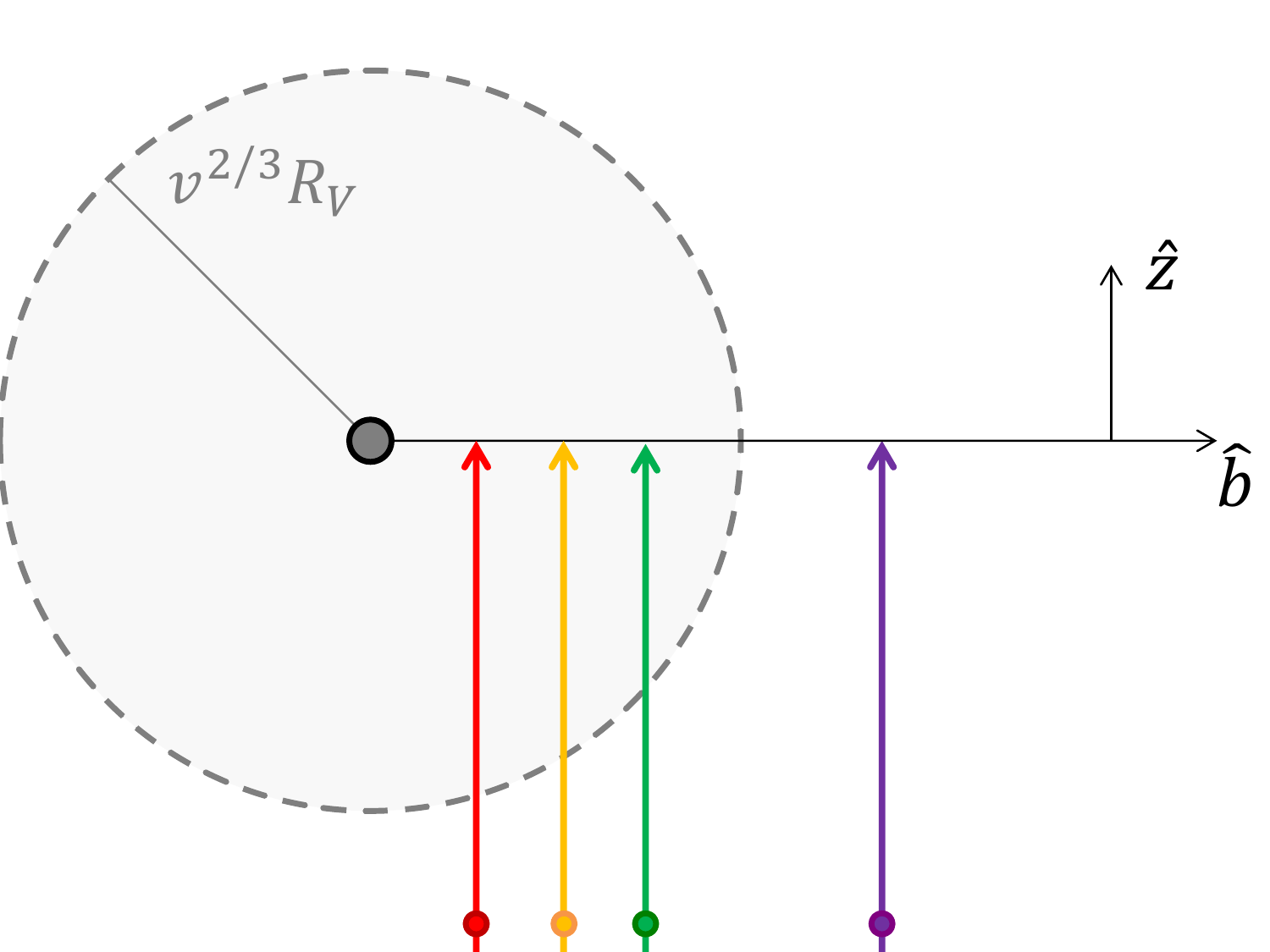}
\includegraphics[width=0.58\textwidth]{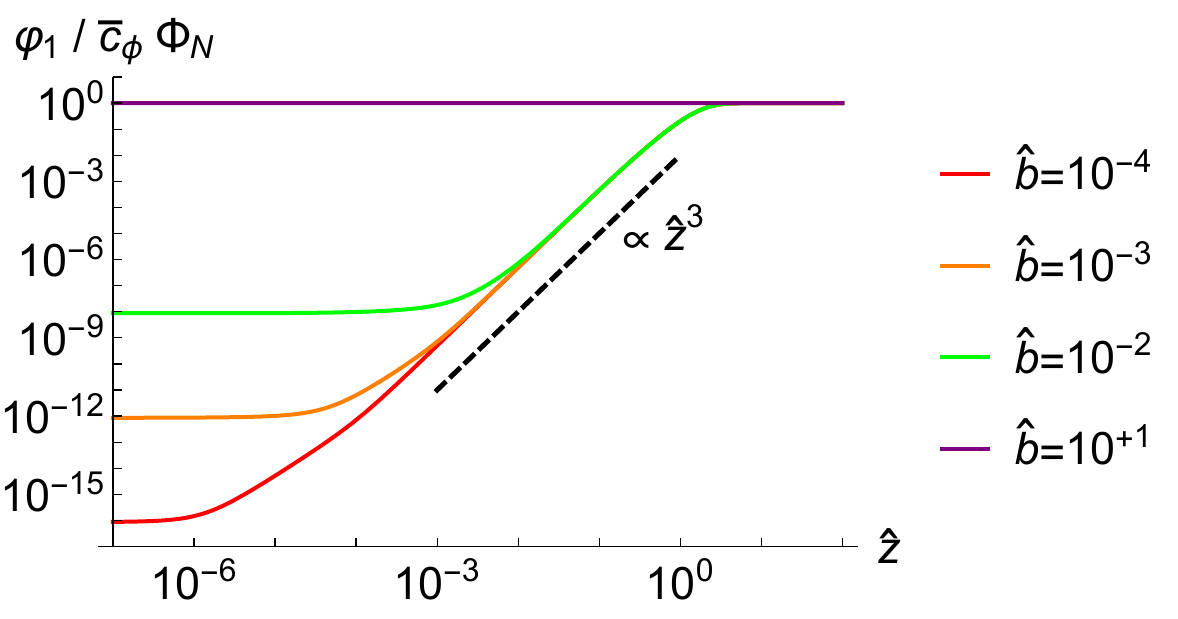}
\caption{
For Rutherford scattering (a particle travelling at constant speed $v$ past another particle at rest, with impact parameter $b$), there is a region around the static particle of radius $v^{2/3} R_V$ within which the disformal effects lead to screening. Above we show four such particle trajectories, with impact parameters $\b = 10^{-4}, \, 10^{-3} , \, 10^{-2}$ and $10$, in units of the screening scale. The dimensionless field profile $\varphi_1 / M_P$ experienced by particle 2, relative to the Newtonian potential $\Phi_N = - R_S/r$, is plotted against the dimensionless $\z$. 
\label{fig:flyby}}
\end{figure}

As an example, for the particular values\footnote{
Note that it is only possible for both $a_1$ and $a_2$ to vanish if $\dis$ has the correct overall sign, since $a_2$ depends on $1+\sqrt{\nu}$, which is either 0 or 2.  
} $\nu=1$, $\dis < 0$ and $\bar{c}_1 \ll 1$ (i.e. $a_1=0,  \, a_2 = 0  , \,  a_3= -1/2 , \, a_4 \approx 0.43  $), this gives a solution,
\begin{equation}
 G_1 = \begin{cases} 
        1 + ...  \;\;\; 							&\z \gg 1  \\
	 \frac{1}{2} \z^3 + ...						&\b \ll \z \ll 1  \\
	\sqrt{\b^2 + \z^2} \left( 2 a_4 \b^3 + ...  \right)		& \z \ll \b 
       \end{cases}
\end{equation}
which is plotted in Figure~\ref{fig:flyby}. The non-perturbative coefficient $\bar{c}_1$ can in principle be measured by going very close to the source, but there the screening is most efficient and the disformal effects are small.

\subsection{Circular Orbits}

Now we move on to gravitationally bound orbits, for which $v^2 \sim R_S / r$. 
The simplest such solution for the reduced body motion is,
\begin{equation}
 r (t) = a  \;\;\;\; , \;\;\;\; \theta (t) = \sqrt{\frac{R_S'}{a} } \, a t
\end{equation}
which corresponds to circular orbits of radius $a$ around the center of mass.

Now since $\partial_t | \mathbf{x}_{12} | = 0$ in this case, \eqref{eqn:Geqn} simplifies to,
\begin{equation}
 G_1 (t)  -  \dis^2 \nu \, \frac{ R_{V_1}^6 }{ a^4 } \,   \partial_t^4 G_1 (t)  = 1  
\end{equation}
the solution to which (consistent with the ladder expansion at large $a$) is simply $G_1 = 1$. 
What this example demonstrates is that it is \emph{not} the absolute velocity of the particles which appears in the ladder parameter $L$, but rather it is the radial velocity which matters, namely $\partial_t | \mathbf{x}_{12} |$. 

For perfectly circular orbits, the disformal effects are only present at next-to-leading order, for instance when going beyond the Newtonian limit to include the effects of the retarded separations. 
This is an important exercise, since black hole binaries can quickly circularise, but is postponed for future work. 
In practice, most of the orbits we encounter do have a small eccentricity, and this gives a finite ladder parameter $L$. For instance, since the radial velocity of an orbit $v_r^2 \sim  e^2 R_S'/a$, for typical dark energy scales the condition $L \ll 1$ to neglect disformal corrections corresponds to $e^2 \ll 10^{-15}$ for Earth-like orbits around the Sun. The actual $e$ of the Earth's orbit is $e^2 \approx 10^{-4}$, placing us firmly in the $L \gg 1$ regime once the non-zero eccentricity is taken into account.  
We will therefore pass swiftly on to the case of elliptic orbits.

\subsection{Elliptic Orbits}

In this subsection we will solve for the orbital motion of a binary which is bound gravitationally by $\hat{\mathbf{F}}_h$, and hence has $v^2 \sim R_S/r$. 
First we will review the leading order solution (the well-known Keplerian elliptic orbits), and then discuss how this is modified in the ladder expansion regime, $r^4 \gg  R_S R_V^3$, and in the ladder resummation regime, $r^4 \ll R_S R_V^3$.

\paragraph{Keplerian Orbits:}
The leading order geodesic equation \eqref{eqn:Eeom} has solution,
\begin{equation}
 r (t) = \frac{ a (1-e^2) }{ 1 + e \, \cos \theta (t) }
 \label{eqn:rKepler}
\end{equation}
where $a$ is the semimajor axis determining the size of the orbit, and $e$ is the orbital eccentricity, related to the previous constants of motion by,
\begin{equation}
 J = \sqrt{ R_S' \, a (1-e^2) } \;\; , \;\;\;\; E = - \frac{R_S'}{2 a}  \, . 
\end{equation}
The eccentricity, $e$, obeys $0 \leq e < 1$ for bound orbits, where $e \to 0$ corresponds to the circular limit (constant $r$). 

The range of spatial separations is,
\begin{equation}
 r_- \leq r \leq r_+ \;\;\;\; \text{where} \;\; r_{\pm } = a ( 1 \pm e ) \, , 
\end{equation}
and the range of radial velocities is,
\begin{equation}
 0 \leq  \dot{r}^2 \leq   \frac{e^2}{ 1-e^2 } \,\frac{R_S}{a} \,  ,
\end{equation}
where the maximum is achieved when $r = a (1-e^2) $, and for small $e$ is suppressed by $e^2$ relative to the naive $v^2 \sim R_S/a$ expectation (since circular orbits have no change in $r$). 
The ladder parameter in this case is $L = e^2 R_{S}' R_{V_1}^3 / a^4$.
We will quote the resuls in this section to leading order in small $e$, 
which should be understood as $1 \gg e \gg a^4/R_S R_V^3$ when in the ladder resummation regime $ L \gg 1 $.

\paragraph{Ladder Expansion:}
In the ladder expansion regime, we have a correction to the scalar field, 
\begin{align}
 G_1 (t) 
 &= 1 +  \dis \,  R_{V_2}^3  \partial_t^2 \left(  \frac{1}{r (t)}  \right)    
\end{align}
where $r(t)$ is given by the Newtonian solution \eqref{eqn:rKepler} at leading order. 
This provides a correction to the conformal fifth force, 
\begin{equation}
\delta F_{\varphi \, 2}^{(\con) \; r} = -  \frac{ \dis \con^2 }{2 a }  \; \frac{ R_{S_1} R_{V_2}^3 }{ a^4 } \;  
 \frac{ R_S ( 1 + \frac{\con^2}{2} )  }{ a } \; \left(  e \cos \theta + \mathcal{O} ( e^2 ) \right)    \, . 
\end{equation}
We also have the disformal fifth force generated by the leading $\varphi^{(0)}_1$ profile,
\begin{align}
\delta F_{\varphi  \, 2}^{ (\dis)\; r } 
&=  \frac{\dis \con^2}{ a } \, \frac{ R_{S_1} R_{V_1}^3}{  a^4 }  \frac{ R_S ( 1 + \frac{\con^2}{2} )  }{ a }  \;\left( e \cos \theta  + \mathcal{O} ( e^2) \right) \, .
\end{align}
The angular forces, $F_{\varphi \, A}^{\Omega}$, remain zero and so the angular momentum remains conserved.
Note that both of the radial forces are suppressed by one power of $e$, and thus vanish for perfectly circular orbits. Also note that the mass dependence is different: in the limit $m_1 \gg m_2$, it is the disformal fifth force which provides the largest correction to the orbit.

\paragraph{Ladder Resummation:}
Now let us contrast this with the ladder resummation regime, $a^4 \ll R_S' R_V^3$. 
In terms of the eccentric Binet variable\footnote{
Solutions to \eqref{eqn:Geqn} in this regime are most conveniently found in terms of the \emph{eccentric anomaly}, $\E (t)$, which is related to $\theta (t)$ by,
\begin{equation}
 r (t) = a ( 1 - e \cos \, \E (t) )  , \;\;\;\; \cos \theta (t) = \frac{ \cos\, \E (t)  - e }{ 1 - e \cos \, \E (t) } \, ,
\end{equation}
and has the property that $r (t) \dot{\mathcal{E}} (t) =  a \omega $, where $a \omega = \sqrt{R_S'/a}$ is a constant frequency. This allows the time integrals in \eqref{eqn:fnresumPN} to be performed straightforwardly, and of the resulting complementary functions we find that only $f_1$ is periodic. We will retain only $c_1$ as a free parameter, so that the leading order Newtonian solution for the $( \mathbf{x}_{12} , h_{\mu\nu}  , \varphi   )$ system is completely periodic. Although it would be interesting to investigate the possibility of non-perturbative corrections which are non-periodic, to really diagnose whether other $c_n$ are consistent with our asymptotic boundary conditions would require going beyond the $L \gg 1$ approximation made here.  
} $\rho = r (t)/a$,  
the most general periodic solution to \eqref{eqn:Geqn} in this regime is,
\begin{equation}
 G_A =  \frac{ a^4 }{\dis  R_{S_A} R_{V_A}^3 } \; \frac{\rho}{2 + e^2}  \left[   
 c_1     
 +  \frac{ 1 -  e^2  }{2}  \rho^2  
 + \frac{  1  }{ 3 } \rho^3 \right] \, .
\label{eqn:GAPI}
\end{equation}
It contains one undetermined coefficient 
which reflects a potential non-perturbative correction to the ladder expansion. We take $c_1$ to be order one or smaller, to ensure that $G_A \sim  \frac{ a^4 }{\dis  R_{S_A} R_{V_A}^3 }$ overall (otherwise the Newtonian EOB approach breaks down).  

Taking \eqref{eqn:GAPI} as the leading order estimate of $G_A$, then the conformal and disformal fifth forces are the same order, and are given by, 
\begin{align}
 \delta F_{\varphi \, 2}^{(\con) \, r} &= - \frac{ \con^2}{4 \dis a } \frac{ R_{S_2}}{a} \, \frac{ a^4 }{ R_{S_1} R_{V_1}^3 } \;   \left[   
 \frac{ c_1 }{ \rho }    
 +  \frac{ 1 }{2}  \rho  
 + \frac{  1  }{ 3 } \rho^2 \right]     \\
 \delta F_{\varphi \, 2}^{\dis \, r} &= \frac{ \con^2 }{ 48 \dis a } \frac{ R_{S} }{a} \frac{ a^4 }{ R_{S_1} R_{V_1}^3  }   
 \Big[ 
\frac{36 c_1^2}{\rho ^6}-\frac{96 c_1^2}{\rho ^5}+\frac{12 c_1 \left(7
   c_1+5\right)}{\rho ^4}-\frac{24 c_1 \left(c_1+5\right)}{\rho ^3}+\frac{21-4
   c_1}{\rho ^2}  \nonumber  \\
   &\qquad\qquad\qquad\qquad+\frac{4 \left(26 c_1-7\right)}{\rho }-5 \left(8
   c_1+7\right)-\frac{32 \rho ^3}{3}+\frac{40 \rho ^2}{3}+\frac{118 \rho }{3}
   \Big]
\end{align}
at leading order in $e$, and are suppressed like $1/L$ when $L \gg 1$. 

~\\

This concludes our various examples of ladder screening, which together clearly establish that when $L \ll 1$ one can expect (modulo non-perturbative corrections) an efficient screening mechanism to take place. Even very simple theories, such as \eqref{eqn:simpleS}, with a disformally coupled scalar are therefore \emph{not} necessarily ruled out by two-body solar system tests. 
A re-analysis of the existing experimental constraints on disformal theories, taking into account this ladder screening effect, is urgently needed, as is future model-building to embed this mechanism into a scalar field model which realises desired dark matter / dark energy phenomenology.  
We will now describe the planetary precessions in more detail, and use our results to place rudimentary constraints on the $(\con, \dis)$ parameter space (Figure~\ref{fig:exclusion}).

\begin{figure}
\centering
 \includegraphics[width=0.5\textwidth]{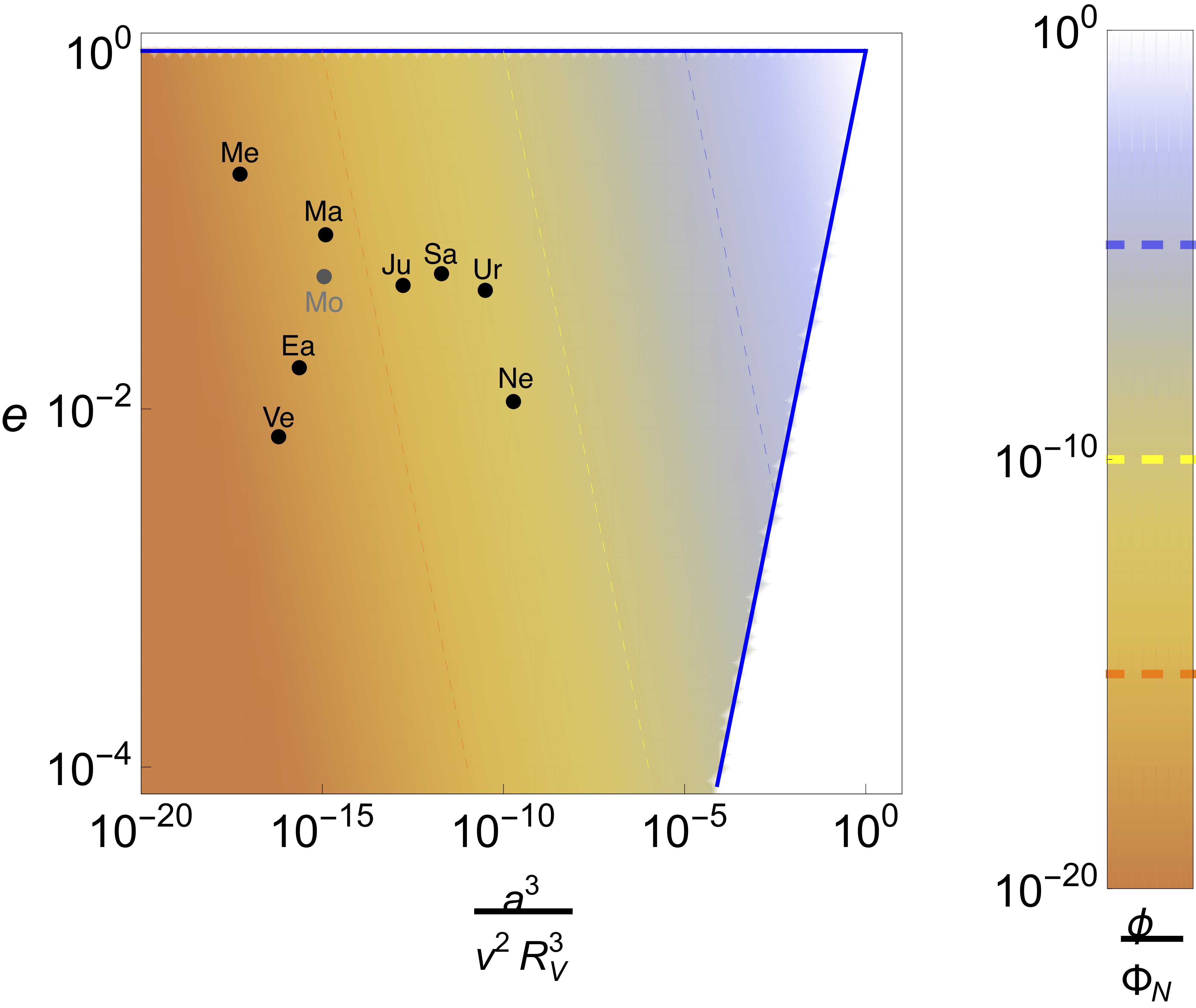}
\caption{With order one couplings ($ \con^2/\dis = 1$), bound orbits in the ladder resummation regime ($1 \gtrsim e \gg a^3/v^2 R_V^3 \sim a^4/R_S R_V^3$) experience a fifth force from a disformally coupled scalar field which is screened by many orders of magnitude relative to the Newtonian potential, $\Phi_N = - R_S/r$. Planets in our solar system are shown as black points, and the Moon is shown as a grey point. The screening is less effective for the outermost planets, as discussed in Section~\ref{sec:solarsystem}. 
}
\label{fig:screening}
\end{figure}

\section{Solar System Constraints from Planetary Ephemerides}
\label{sec:solarsystem}

The perihelion precession of the planets can be used to constrain $\con$ and $\dis$. In this Section we derive such constraints, and ultimately find that (because of the screening effects) they are weaker than previously estimated in the literature.  

\begin{figure}
\includegraphics[width=0.55\textwidth]{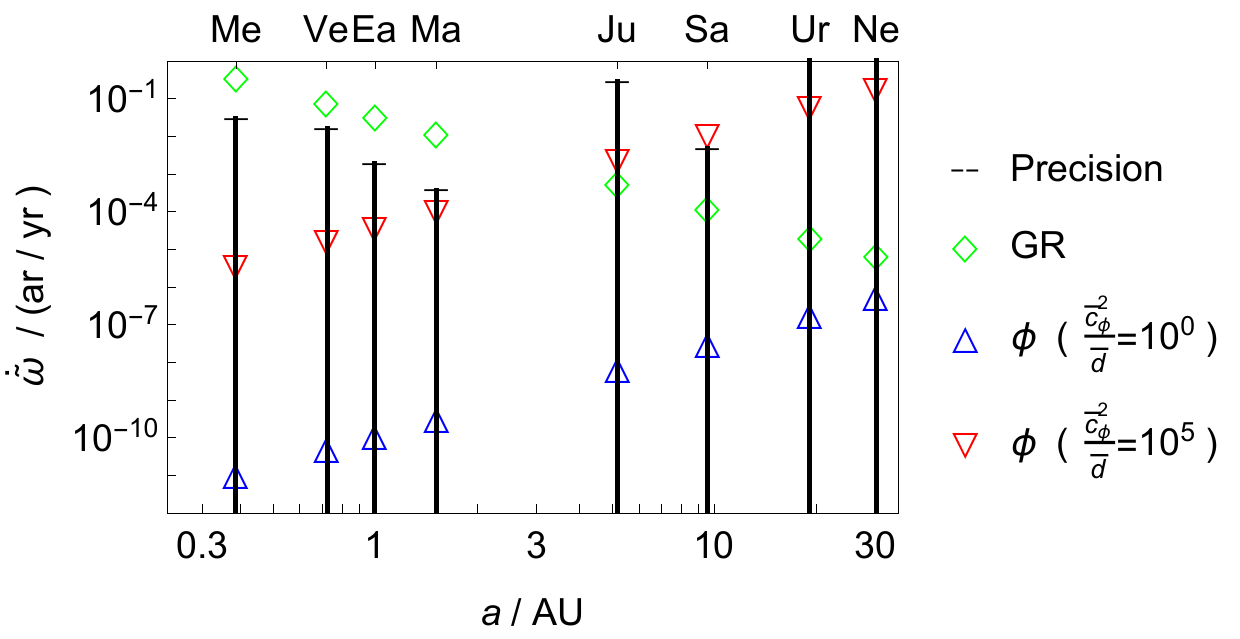}
\includegraphics[width=0.45\textwidth]{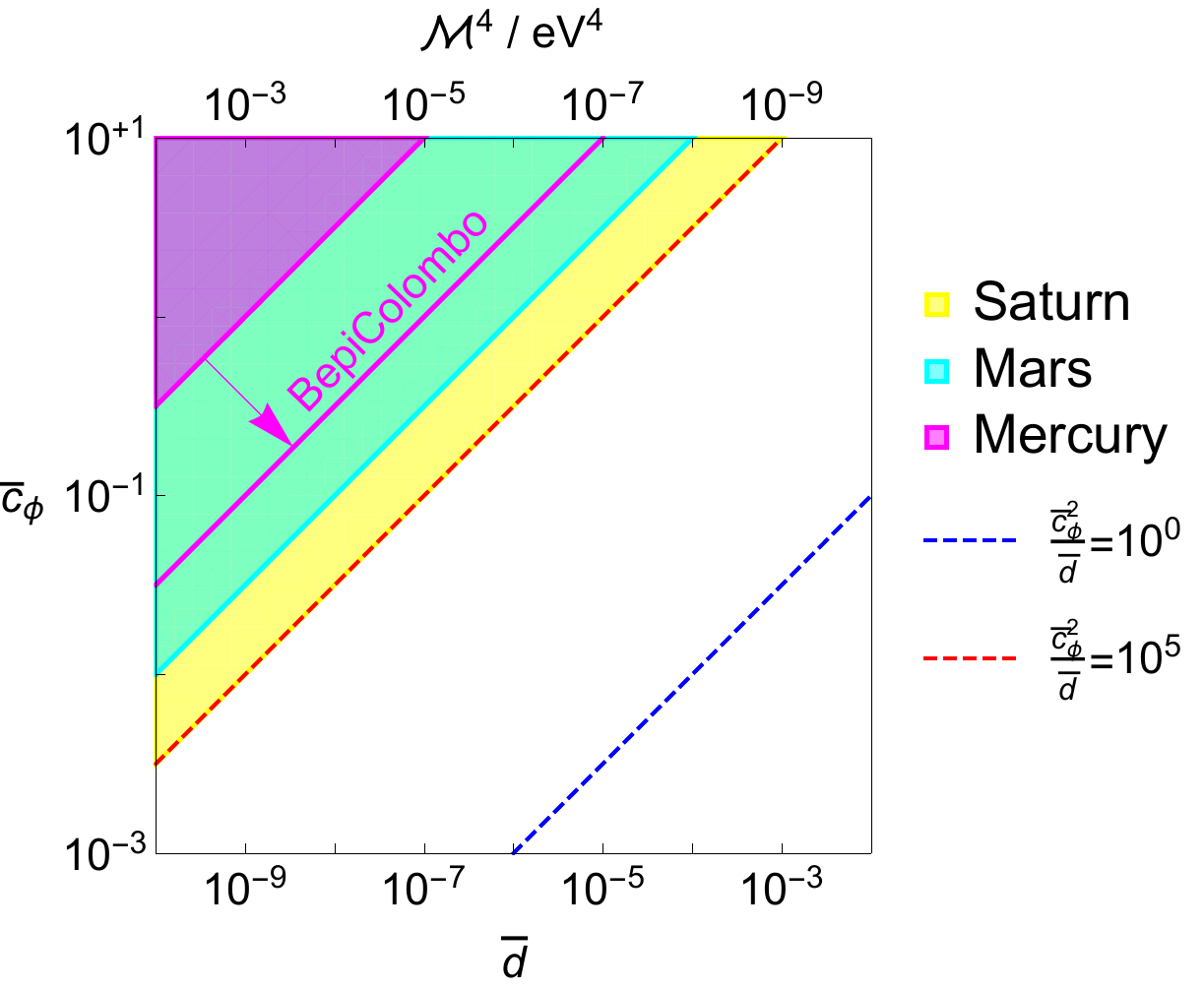}
\caption{The precision with which we can account for orbital precession of each planet in the solar system is shown as a black error bar. 
The relativistic corrections predicted by GR are resolvable for the innermost planets (Mercury, Venus, Earth and Mars). The precession predicted here due a disformally coupled scalar field is shown for two fiducial values of the couplings, $\con^2/\dis=1$ and $\con^2/\dis = 10^5$. 
The ladder screening is less effective for the outermost planets, and they experience the largest disformal effects. Saturn has lower than expected error bars thanks to the \emph{Cassini} measurements, and so as $\con^2/ \dis$ increases it is Saturn's orbit which first reveals the presence of a disformally coupled scalar. The region of $(\con/\dis)$ parameter space ruled out by our observations of Mercury, Mars and Saturn is also shown (i.e. $\con^2/dis = 1$ is not currently resolvable in any planetary orbit, but $\con^2/\dis = 10^5$ would have been observed already in measurements of Saturn and so is ruled out), together with the improvement forecast by the \emph{BepiColombo} mission to Mercury.  
\label{fig:exclusion}
}
\end{figure}

\paragraph{PN Corrections to Orbit:}
To compute the PN corrections to the Newtonian separation $\mathbf{x}_{12} (t)$ from subleading
terms in $g_{\mu \nu}$ and the disformal force from $\varphi$, a more sophisticated effective one-body scheme would be required\footnote{
This can be understood as replacing $m_A \mathbf{v}_A$ with the modified momenta,
\begin{equation}
p_{A \, \mu} ( \tau ) = - \tilde{e}_A \frac{ \partial S_A }{ \partial \dot{x}_A^\mu (\tau) } = m_A  \left(  \eta_{\mu\nu} \dot{x}_A^{\nu} + \frac{h_{\mu\nu} (x_A) }{M_P}  \dot{x}_A^{\nu} + \frac{\dis}{\M^4} \frac{\partial \varphi ( x_A )}{\partial x_A^\mu}  \;  \partial_\tau  \varphi ( x_A )        \right)    \, .
\end{equation}
and accounting for the resulting mixing between $m_1 \mathbf{v}_1 + m_2 \mathbf{v}_2$ and $\mathbf{x}_{12}$ perturbatively.
}, since the subleading forces do not obey \eqref{eqn:EOB} (see e.g. \cite{Damour:2011xga} for a useful review). While it would be interesting to extend existing one-body techniques to this disformally coupled case (and potentially very relevant for comparable-mass black hole/neutron star binaries), we will instead take our computed $G_A (t)$ (valid for any $m_1$ and $m_2$) and focus on the limit\footnote{
As with the velocity and the eccentricity previously, we treat $ \frac{a^4}{ R_{S_1} R_{V_1}^3 } \ll  \nu \ll 1 $ in the ladder resummation regime so that the leading two-body effects are still present even though $\nu$ is small. 
} $\nu \ll 1$, since then we can work entirely in the rest frame of particle 1 (its motion relative to particle 2 is suppressed by powers of $\nu$) and compute the orbital corrections straightforwardly. 
Another simplification in this is that the conformal force $ \delta F_{\varphi \, 2 }^{(\con) \, r}$ is suppressed relative to the disformal force $ \delta F_{\varphi \, 2 }^{(\dis) \, r}$ by a factor of $\nu$, so we need only follow the effects of $\delta F_{\varphi \, 2 }^{(\dis) \, r}$.  
 
The forces induced by the disformal interaction act as a source in the orbital equation, 
\begin{equation}
 \ddot{r_2} - \frac{J^2}{ r_2^3 } =  - \frac{R_S'}{ r_2^2}   +  \delta F_{\varphi  \, 2}^{ (\dis)\; r } ( r_2 )   + ... \,\, . 
\end{equation}
where $r_2 = | \mathbf{x}_2 | - R_S' $ is the effective radial variable for $\mathbf{x}_2 (t)$ in the rest frame of particle 1.
In terms of the Binet variable, $u (\theta) = a (1-e^2)/ r_2 $, we can expand about the Newtonian solution $u = u_N + \delta u + .. $, where,
\begin{equation}
\delta u'' (\theta ) + \delta u (\theta) =  - \frac{1}{u_N^2 (\theta)}  \frac{ \delta F_{\varphi  \, 2}^{ (\dis)\; r } ( u_N (\theta) )}{ R_S' / a }  \, . 
\end{equation}
At small $e$, the right hand side can always be expanded as a series of $\cos ( n \theta )$ forcing terms. This sources many higher harmonics in $\delta u$, but also terms which are non-periodic,
\begin{equation}
 \delta u ( \theta ) \supset  - \alpha \; e \,  \theta \sin \theta
 \label{eqn:uprecess}
\end{equation}
and which contribute over many cycles to a growing precession of the orbit. 
Explicitly, after each orbit the perihelion precesses by,
\begin{equation}
 \Delta \theta = 2 \pi  \alpha  
\end{equation}
in radians per orbit, or,
\begin{equation}
\dot{ \tilde{\omega} } 
= 60^2  \alpha \,  \frac{180}{2 \pi a} \, \sqrt{ \frac{ R_{S}' }{ a } }
\end{equation}
in arc seconds per unit time (using the fact that the orbital period $T = 2 \pi / \omega = 2 \pi a \sqrt{a/R_S'}$).

\paragraph{Precession in Ladder Expansion:}
The forces induced by the disformal interaction lead to an orbital precession, governed by,
\begin{equation}
\delta u'' (\theta ) + \delta u (\theta) = -\dis \con^2  \frac{ R_{S_1} R_{V_1}^3 }{ a^4}   \;\left(  e \, \cos \theta  + \mathcal{O} (e^2)  \right)
\end{equation}
$\delta u (\theta)$ contains harmonics $\cos (n \theta)$ up to $n=5$, but also a non-periodic piece \eqref{eqn:uprecess}, which leads to a precession\footnote{
Without treating $e$ as small, the precession would be given by, $ \alpha = \frac{ R_{S_1} R_{V_1}^3 }{ a^4}  \; \frac{ 4 - 35 e^2 - 110 e^4 - 10 e^6 }{8 (1-e^2)^3}$.
},
\begin{equation}
 \Delta \theta = 2 \pi  \alpha  \;\;\;\; \text{with} \;\;\;\; \alpha = \frac{\dis \con^2 }{2} \; \frac{ R_{S_1} R_{V_1}^3 }{ a^4}  
\label{eqn:ladder_precession}
 \end{equation}
in radians per orbit. 

This result is of most interest in the following regime, 
\begin{equation}
 \left( R_S^3 R_V^3 \right)^{1/6}  \gg  a \gg  \left( R_S R_V^3 \right)^{1/4} \, , 
\end{equation}
where the orbit is big enough to allow for the ladder expansion, but small enough that the disformal correction dominates over the usual GR one. This is particularly relevant if $\phi$ is to represent a UV-modification of gravity which becomes important in the strong field regime, or even perhaps if $\phi$ were to play the role of dark matter, since then $\M$ may be sufficiently high that this regime include planets in our solar system. For typical dark energy scales, we are not in this window and we instead require a ladder resummation to describe planetary motion.

\paragraph{Precession with Ladder Resummation:}
At small $e$, the radial component of the disformal force felt by particle 2
is dominated by,
\begin{equation}
\delta u'' (\theta ) + \delta u (\theta)  = - \frac{ \con^2}{\dis} \, \frac{a^4}{ R_{S_1} R_{V_1}^3}  \, \frac{ 5 + 6 c_1 }{12} \, \left( e \cos \theta + \mathcal{O} (e^2)  \right)   \;\; ,  
\end{equation}
which gives a precession\footnote{
Without treating $e$ as small, the full disformal force is,
\begin{align}
 \mathbf{F}_{\varphi \, 2}^{(\dis) \, r} =& \frac{ \con^2 }{\dis a } \frac{ R_{S} }{a} \frac{ a^4 }{ R_{S_1} R_{V_1}^3  }  \frac{ 1 }{ ( 2 + e^2 ) \rho^3}  \left[   
 c_1   
 +  \frac{ 1 -  e^2  }{2}  \rho^2  
 + \frac{  1  }{ 3 } \rho^3 \right]   \nonumber \\
 & \times \; \frac{1}{12 (2+e^2) \rho^3} \left[ 42 \rho^2 - 84 \rho^3 - 14 \rho^4 + 88 \rho^5 - 32 \rho^6 + c_1 \left(  36 - 96 \rho + 84 \rho^2 - 24 \rho^3    \right)        \right] \, . 
\end{align}
It is difficult to find the orbital precession from such a force at general $e$ since there are $1/u$ terms in the Binet equation, but at small $e$ we can expand as in the main text, and arrive at  \eqref{eqn:PrecessionSmalle} for the estimated orbital precession due to disformal effects. 
},
\begin{equation}
 \Delta \theta = 2 \pi  \alpha  \;\;\;\; \text{with} \;\;\;\; \alpha = \frac{ \con^2 }{ \dis } \, \frac{ 5 + 6 c_1 }{24}  \, \frac{a^4 }{  R_{S_1} R_{V_1}^3 }
 \label{eqn:PrecessionSmalle}
\end{equation}
in radians per orbit.

Contrast this with \eqref{eqn:ladder_precession}. Had we naively taken the ladder expansion result, and applied it for small orbits, with $a^4 \lesssim R_S R_V^3$ (such as the planets in our solar system), we would have incorrectly concluded that the disformal scalar leads to very large effects. Rather, we see that when orbital separations are sufficiently small (and hence orbital velocities sufficiently large), that the effect of the ladder resummation is to \emph{screen} the effects of the scalar and restore approximately Newtonian orbits, with corrections of $\mathcal{O} \left( a^4 / R_S R_V^3   \right)$.  
The magnitude of this suppression for the orbits in our solar system are shown in Figure~\ref{fig:screening}.

\paragraph{Solar System Constraints:}
We have shown that the effects of such a screened fifth force result in a precession of planetary orbits. If no anomalous precession is detected in our observations (i.e. no precession in addition to what can be explained by GR and existing solar system astronomy), then this can be used to place constraints on the ($\con, \, \dis$) parameter space of our disformally coupled toy model\footnote{
There are other ways to constrain screened fifth forces (see for instance \cite{Sakstein:2017pqi}), but here we will focus on orbital precession for concreteness. 
}. In the following discussion, we will assume typical dark energy scales $\M \sim M_P H_0$ for the disformal interaction.   

Currently, some of our best measurements of Mercury's orbit are due to the \emph{MESSENGER} mission (which orbited Mercury from 2011 until 2015) \cite{Fienga:2011qh,Verma:2013ata,Fienga:2015xpw,Park_2017}, and placed experimental bounds on the perihelion precession on the order of $10^{-3}$ ar/yr.   
An ongoing mission, \emph{BepiColombo} \cite{BENKHOFF20102}, launched in 2018 and due to orbit Mercury from 2025, has the potential to improve this accuracy by an order of magnitude \cite{Milani:2002hw, Ashby:2007kw}, down to approximately $10^{-4}$ ar/yr.  
At such accuracies, \emph{BepiColombo} may resolve novel General Relativistic effects \cite{Will:2018mcj}, and would also detect a precession from the disformal force given above if $\con^2/\dis > 10^{7}$. 
  
Estimating the orbital precession for the other planets in the solar system is no mean feat, since they are affected by a variety of astrophysical effects (Newtonian forces from other planets and their moons, solar oblateness, asteroids, etc.).
However, progress was made in \cite{2005SoSyR..39..176P, 2005AstL...31..340P, 10.1093/mnras/stt695} by processing data from 1913-2011 and analysing the EPM2011 ephemerides \cite{Pitjeva:2013fja}---in particular \cite{10.1093/mnras/stt695} provides an estimate of the observational bounds on any anomalous precession in the planetary orbits from Mercury out to Saturn (see also \cite{Fienga:2016ynj} for similar estimates for Mercury and Saturn, and further discussion in \cite{Iorio:2005fk, Iorio:2008sd}. Earlier estimates are also given in \cite{standish1992orbital, fitzpatrick2012introduction}). 
Interestingly, the error bars for Saturn are lower than may have been expected thanks to \emph{Cassini} measurements, and we would have already detected a precession due to the above disformal force if $\con^2/\dis > 10^{5}$.

We display these experimental precisions as error bars in Figure~\ref{fig:exclusion}, together with the precession from GR corrections and the predicted effects of a disformally coupled scalar (for two fiducial values of $\con^2/\dis$). If the predicted effects from fifth forces are \emph{larger} than our current experimental precision, then this value of ($\con,\dis$) should be excluded.
Since the screening is less efficient for the outermost planets, it is the orbit of Saturn which is most sensitive to disformal effects and rules out the largest fraction of parameter space. Even with the significant improvement that \emph{BepiColombo} will provide on our measurements of Mercury's orbit, since the screening is so strong there it will not rule out any theories which are not already ruled out by Saturn (in this simple setup \eqref{eqn:simpleS}).

\section{Discussion}
\label{sec:disc}

We have shown that disformally coupled scalar fields have a regime, in which separations are small and relative velocities are large, that leads to the efficient screening of fifth forces. 
In this regime, the conventional ladder expansion breaks down, and must be replaced by a complete resummation of ladder diagrams---we have achieved this in the form of a functional differential equation, and shown that (at least for a head-on collision in the Newtonian limit) this agrees with other resummation techniques (namely Borel resummation). On the theoretical side, we have also shown that this resummation can introduce additional free parameters into the theory, capturing the possible non-perturbative effects which are invisible in perturbation theory.    
On the phenomenological side, we have shown that our ladder screening mechanism can render even the simplest disformally coupled scalar compatible with two-body solar system tests. 

We believe that this two-body mechanism is distinct from the usual Vainshtein mechanism, and also quite different from any previously proposed ``disformal screening'' mechanism. 
In an upcoming work \cite{us}, we will describe in detail how this theory in the Jordan-frame (the quartic DBI Galileon) does not exhibit conventional Vainshtein screening (due to an interesting cancellation of all operators at $\Lambda_3$), and appears consistent from the point of view of radiative stability and positivity bounds.  
 
Building from these results, there are a number of interesting directions to pursue in future. 
For instance, for concreteness we have focussed on the orbital precession which would be induced by such a screened disformal force, however there are numerous other experimental probes of such disformal couplings, and it is now vital that they are also revisited with the potential for ladder screening in mind.    

Furthermore, to exemplify the ladder resummation effects in the simplest possible setting, we have mostly considered a free scalar field with constant conformal and disformal couplings to matter. Of course, a more sophisticated scalar-tensor sector may be desired if $\phi$ is to describe dark energy, dark matter, or some strong gravity effect from new UV physics. 
In particular, although we have focussed somewhat on the scales relevant for dark energy, it would be interesting to incorporate this screening mechanism into a dark matter model, since this screening effect naturally gives rise to a modified force law inside galactic haloes. 
Moreover, if one imposes that the speed of tensor modes is approximately equal to that of light---motivated\footnote{
though not required in the case of dark energy, as details of the UV completion are required to assess how much $c_{GW}$ runs as one approaches LIGO scales \cite{deRham:2018red}. In particular, since the dependence of $c_{GW}$ on $k$ near the cutoff is completely unknown within the EFT, it is not possible to estimate how far the cutoff must be below $\Lambda_3$.     
} by the GW170817 multi-messenger observation---then the disformal coupling must be very small, and it would be interesting if future model-building could employ a similar ladder resummation to achieve screening in a way compatible with luminal tensor modes.  

Such a resummation of ladder diagrams in the two-body problem can occur in a wide variety of contexts, not just the disformal coupling in a scalar-tensor theory. 
For instance, an analogous resummation leads to the Sommerfeld enhancement effect at low relative velocities and the Bethe-Salpeter equation for the formation of bound states in a relativistic field theory. The key difference for the disformal coupling considered here is that, while the vertex interaction is usually a differential operator, in this point particle case it is an integral operator $\sim \int d\tau \; \delta ( x - x_A(\tau)  )$, which results in a \emph{functional} relation between the scalar field at different times. 
Note also that we have neglected the subleading conformal term, $ \bar{c}_X (\nabla \phi)^2  / \M^4$ in the effective metric $\tilde{g}_{\mu\nu}$, and also the self-energy terms in $\varphi_1 T_1$ and $\varphi_2 T_2$. Given the novel (and unexpected!) behaviour we have found in our simple disformally coupled theory, it would be interesting in future to revisit both of these omissions (particularly in cases where the ladder screening yields a $\varphi_A$ which is \emph{finite} on the worldline $x_A$).  

Also, extending the above discussion to a cosmological background---with respect to which even single particles can have a relative velocity $v^2$ and hence a non-zero ladder parameter---may produce interesting effects such as screening (for instance the Milky Way has a velocity $v/c \approx 10^{-3}$ with respect to the CMB rest frame, which can lead to large preferred frame effects \cite{Ip:2015qsa}).  
 
Finally, although the binary system we have considered here is phenomenologically very relevant for black hole/neutron star signals, it would be interesting to investigate the effects of multiple bodies. 
$N$-body effects are relevant for many solar system observations, and can allow for novel tests of modified gravity\footnote{For instance, observations of the Earth-Sun-Jupiter system could provide tests of the strong equivalence principle with order $10^{-15}$ precision \cite{Anderson:1995df}.}. 

Going forwards, this ladder resummation we have studied will play an important role in the phenomenological viability and testing of scalar-tensor theories that contain a disformal coupling to matter.

\vspace{-0.1in}
\section*{Acknowledgements}
\vspace{-0.1in}
\noindent We thank P.~Brax, A.~Kuntz, J.~Noller, J.~Sakstein and C.~de Rham for discussion and helpful comments. 
SM is supported by an Emmanuel College Research Fellowship.  
This work is supported in part by STFC under grants ST/L000385, ST/L000636 and
ST/P000681/1.

\appendix
\section{Computing the Ladder Expansion Coefficients}
\label{sec:ladder_explicit}

In this short appendix, we will explicitly solve the scalar equations \eqref{eqn:ladder0} and \eqref{eqn:ladder1} for the quoted solutions $\varphi_1^{(0)}$ \eqref{eqn:ladder0sol} and $\varphi_1^{(1)}$\eqref{eqn:ladder1sol}, then the metric equation \eqref{eqn:heom2} for the quoted $h^{\mu\nu}$ \eqref{eqn:eom_metric_fluc}, and finally outline an efficient route to the $\partial_{\tau}$ of these solutions.   

\paragraph{Solving the scalar equation:}
We will employ the retarded Green's function, 
\begin{equation}
 \Box G_{\Box} (x, x') = \delta^{(4)} ( x - x' )  \;\;\;\; \Rightarrow \;\;\;\;  G_{\Box} (x, x') = - \frac{1}{2 \pi} \Theta ( x^0 - x'^0 ) \, \delta \left(  (x-x')^2 \right)
\end{equation}
which propagates signals along the future lightcone only. 

\eqref{eqn:ladder0} is straightforward,
\begin{align}
 \varphi_1^{(0)} (x) &= - \frac{ \con }{2 M_P} \int d^4 x' \, G_{\Box} ( x, x')  T_1 (x)   \nonumber \\
 &= \frac{ \con m_1}{4 \pi M_P} \int d^4 x' \Theta ( x^0 - x'^0 ) \, \delta \left(  (x-x')^2 \right) \int d \tau \,e_1 \,  \delta^{(4)} ( x - x_A (\tau ) )  \nonumber \\
 &= \frac{ \con m_1}{4 \pi M_P}  \int d \tau \, e_1  \; \Theta ( x^0 - x_A^0 (\tau) ) \, \delta \left(  ( x-x_A (\tau) )^2 \right)  \nonumber \\
 &= - \frac{ \con m_1}{4 \pi M_P} \; \frac{e_1}{2 (x-x_A)\cdot u_A } \, \Big|_{\rtau_1}  \nonumber \\
 &= - \frac{ \con R_{S_1} }{ \bar{R}_1 (x) } \, . 
 \label{eqn:ladderworking}
\end{align}

\eqref{eqn:ladder1} is solved in a similar fashion,
\begin{align}
 \varphi_1^{(1)} (x) &= \frac{ \dis }{\M^4} \int d^4 x' \, G_{\Box} (x,x') \, \nabla_\mu' \left(  \nabla_\nu' \, \varphi_2^{(0)} (x') \, T_1^{\mu \nu} (x') \right)
\end{align}
Note that,
\begin{align}
\frac{ \nabla_\mu T_A^{\mu \nu} }{m_A}  =  - \int \frac{d \tau}{e_A}  \, \frac{\partial}{\partial \tau} \delta^{(4)} ( x - x_A (\tau) )  \, u_A^{\nu} ( \tau ) = \int \frac{d \tau}{e_A} \; \delta^{(4)} ( x - x_A (\tau) ) \, a_A^{\nu} ( \tau )
\end{align}
and so, 
\begin{align}
 \int d^4 x' \, G_{\Box} ( x, x' ) \nabla_{\mu} \left[ \nabla_\nu \varphi_A  T_B^{\mu \nu}  \right] &= - \frac{1}{2 \pi} \int d \tau \, \delta ( (x - x_B (\tau) )^2 ) \, \left[  ( u_B (\tau) \cdot \nabla )^2 + a_B (\tau) \cdot \nabla  \right] \varphi_A ( x_B )  \nonumber \\
 &= - \frac{1}{2 \pi} \int d \tau \, \delta ( r_B^2 ) \, \partial_{\tau}^2 \varphi_A ( x_B (\tau) ) \nonumber  \\
 &=  \frac{1}{4 \pi \,  \bar{r}_B \cdot \bar{u}_B }  \partial_{\rtau_B}^2 \varphi_A ( \bar{x}_B  )
\end{align}
yielding, 
\begin{align}
 \varphi_1^{(1)} (x) &= \frac{\dis \, R_{V_1}^3 }{ \bar{R}_1 } \frac{ \partial_{\rtau_1}^2}{ \tilde{e}_1^2 } \, \varphi_2^{(0)} ( \bar{x}_1  ) \, . 
\end{align}
as presented in \eqref{eqn:ladder1sol}.

\paragraph{Solving the metric equation:}
The derivation of the metric solution \eqref{eqn:eom_metric_fluc} proceeds in the same way, where the Green's function now contains a factor of the metric propagator in de Donder gauge on a flat background,
\begin{equation}
\frac{1}{2} \left(   \eta_{\mu (\alpha } \eta_{\beta) \nu} - \eta_{\mu\nu} \eta_{\alpha\beta}   \right) \, . 
\label{eqn:hpropagator}
\end{equation}
The derivation of \eqref{eqn:eom_metric_fluc} from \eqref{eqn:heom2} then proceeds analogously to \eqref{eqn:ladderworking}, with the additional tensor structure \eqref{eqn:hpropagator}. 

\paragraph{Rest Frame Projection:}
At every $\tau$, each particle defines a privileged rest frame, in which $u_A^\mu = ( 1, 0,0,0)$. We can describe this covariantly using the projection tensors,
\begin{equation}
P^A_{\mu \nu} ( \tau ) = g_{\mu \nu} ( x_A ) + \frac{ u_{A\, \mu} u_{A\, \nu}  }{ e_A^2 }
\end{equation}
which project any four-vector onto its spatial components measured in the (instantaneous) rest frame of particle $A$ (at proper time $\tau$). 
This allows us to define the \emph{relative} motion of the two particles via,  
\begin{equation}
u_{AB}^\mu ( \tau) = g^{\mu \nu} (x_A) P^B_{\nu \alpha} u_A^\alpha , \;\;\;\;  a_{AB}^\mu ( \tau ) = g^{\mu \nu} (x_A) P^B_{\nu \alpha} a_A^\alpha
\label{eqn:uAB}
\end{equation}
which give relativistic measures of the ``relative velocity'' and ``relative acceleration'' between the two particles---namely the spatial parts of $u_A^\mu$ and $a_A^\mu$ measured in the rest frame of particle $B$. 
Furthermore, this lets us define a spatial distance from each particle,
\begin{align}
 R_A (x, \tau) = \sqrt{ r_A^\mu P_{\mu \nu}^A r_A^{\nu} } \, . 
\end{align}
which coincides with the earlier definition \eqref{eqn:RAdef} when evaluated at a retarded time.

This notation makes it easier to carry out the $\rtau$ derivatives. For instance, 
\begin{align}
 \partial_{\rtau_1} \varphi_2^{(0)} ( \bar{x}_1 )  &=  \partial_{\rtau_1} \left[   - \frac{ \con R_{S_2} }{ \bar{\bar{R}}_2 }   \right]   \nonumber \\
 &= - \frac{ \con R_{S_2}}{ \bar{\bar{R}}_2^3 } \, \left[ \bar{u}_2^{\mu} \bar{\bar{P}}^1_{\mu \nu} \bar{\bar{r}}_1^\nu + \frac{ \bar{\bar{r}}_1 \cdot \bar{\bar{u}}_1 \, \bar{\bar{r}}_1 \cdot \bar{\bar{a}}_1  }{e_A^2}    \right]  
\end{align}
and so on, where the projection tensors always collect into powers of the relative $u_{12}$, $a_{12}$, etc., evaluated at an iterated retarded time.
Finally, the PN expansion of \eqref{eqn:uAB} is straightforward, allowing an easy connection to the results \eqref{eqn:phiPN} of \cite{Brax:2018bow, Brax:2019tcy}.

\bibliographystyle{apsrev4-1}
\bibliography{Disformal_References}

\end{document}